\documentclass[12pt]{article}

\usepackage{cite}
\usepackage{amsmath}
\usepackage{amssymb}
\usepackage{bm}
\usepackage{color}
\usepackage{graphicx}
\numberwithin{equation}{section}

\newcounter{aff}

\begin{document}
\begin{titlepage}
\begin{flushright}
{\footnotesize NITEP 136, OCU-PHYS 561}
\end{flushright}
\begin{center}
{\Large\bf
Duality Cascades and Parallelotopes
}\\
\bigskip\bigskip
{\large
Tomohiro Furukawa\footnote{\tt furukawatmhr@gmail.com},
Sanefumi Moriyama\footnote{\tt moriyama@omu.ac.jp},
Hikaru Sasaki\footnote{\tt m21sa013@st.osaka-cu.ac.jp}
}\\
\bigskip
${}^{*\dagger\ddagger}$\,{\it Department of Physics, Graduate School of Science,}\\
{\it Osaka Metropolitan University, Sumiyoshi-ku, Osaka 558-8585, Japan}\\[3pt]
${}^{\dagger}$\,{\it Nambu Yoichiro Institute of Theoretical and Experimental Physics (NITEP),}\\
{\it Osaka Metropolitan University, Sumiyoshi-ku, Osaka 558-8585, Japan}\\[3pt]
${}^\dagger$\,{\it Osaka Central Advanced Mathematical Institute (OCAMI),}\\
{\it Osaka Metropolitan University, Sumiyoshi-ku, Osaka 558-8585, Japan}
\end{center}

\begin{abstract}
Duality cascades are a series of duality transformations in field theories, which can be realized as the Hanany-Witten transitions in brane configurations on a circle.
In the setup of the ABJM theory and its generalizations, from the physical requirement that duality cascades always end and the final destination depends only on the initial brane configuration, we propose that the fundamental domain of supersymmetric brane configurations in duality cascades can tile the whole parameter space of relative ranks by translations, hence is a parallelotope.
We provide our arguments for the proposal.
\end{abstract}
\begin{center}
\includegraphics[scale=0.5,angle=-90]{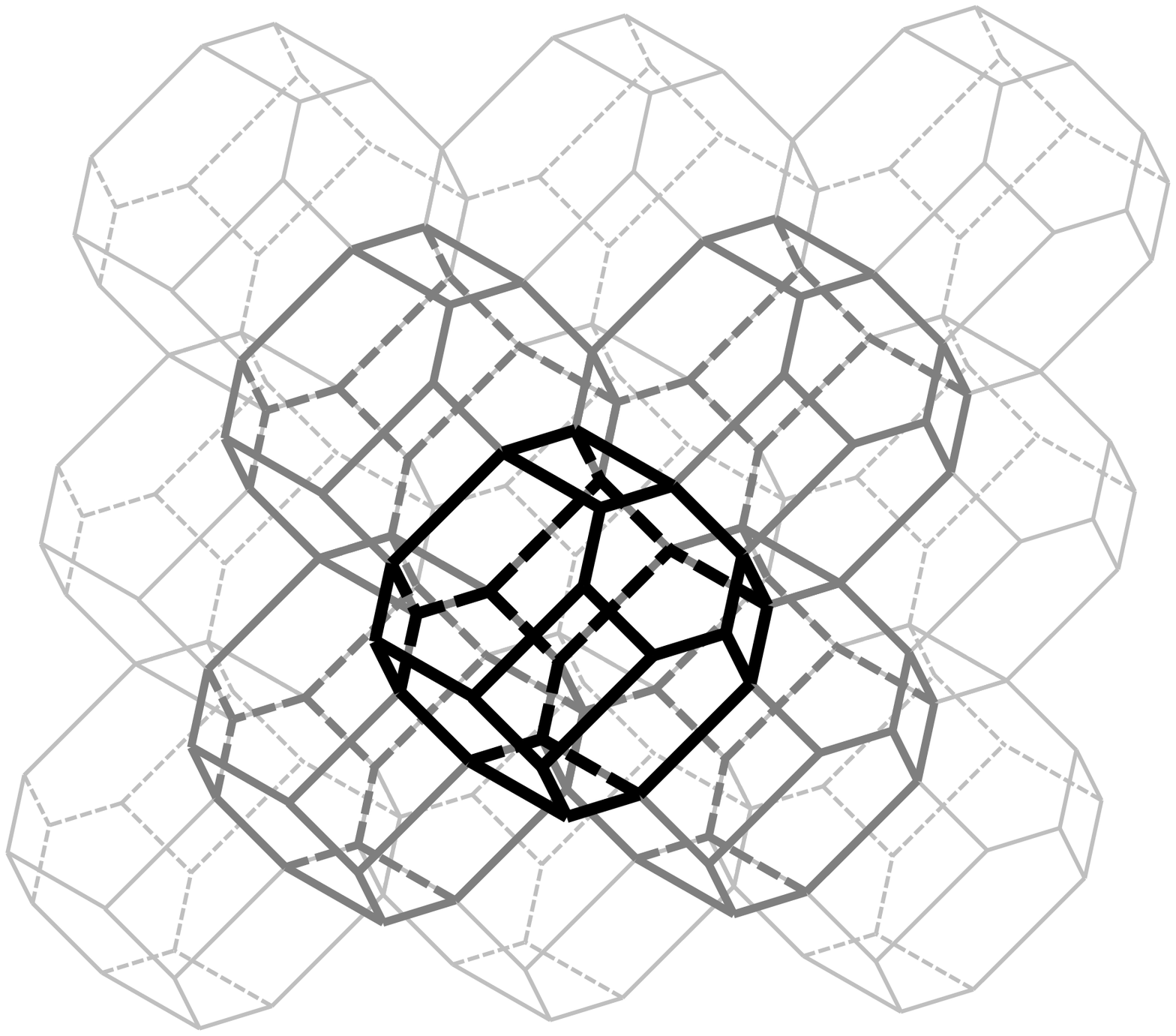}
\end{center}

\end{titlepage}

\tableofcontents

\section{Introduction}

Duality cascades are a series of duality transformations in field theories \cite{KS,SDC}.
Each step of duality transformations is known as the Seiberg duality (and its generalizations) and realized as the Hanany-Witten (HW) transition \cite{HW} in brane configurations on a circle.
Parallelotopes are polytopes which tile the entire space by translations, such as regular hexagons or cubes.
In this paper, we propose deep connections between these two topics.

To explain our main ideas, let us consider the brane configuration for the Aharony-Bergman-Jafferis-Maldacena (ABJM) theory.
The ABJM theory \cite{ABJM,HLLLP2,ABJ} is the three-dimensional ${\cal N}=6$ supersymmetric Chern-Simons theory with gauge group $\text{U}(N_1)_k\times\text{U}(N_2)_{-k}$ and two pairs of bifundamental matters, where the subscripts $(k,-k)$ denote the Chern-Simons levels ($k>0$).
It is known that the theory describes the multiple M2-brane system with $\min(N_1,N_2)$ M2-branes and $|N_2-N_1|$ fractional M2-branes on the background ${\mathbb C}^4/{\mathbb Z}_k$.
The description is obtained from the brane configuration of D3-branes in IIB string theory on a circle with an NS5-brane and a $(1,k)$5-brane located perpendicularly to the circle and tilted relatively to preserve supersymmetries.
Brane configurations on a circle with more numbers of the two types of 5-branes, and hence, more ranks, preserve supersymmetries ${\cal N}=4$ \cite{GW}.
These describe M2-branes on more complicated backgrounds \cite{IK}.
In studying these theories, the main character is the grand canonical partition function $\Xi_{k,{\bm M}}(z)$ which is defined by summing over the overall rank $N$ while keeping the relative ranks ${\bm M}$ fixed (see \cite{DMP1,DMP2,FHM,MP,HMO1,PY,HMO2,CM,HMO3,HMMO,MM,HO} for the ABJM theory and \cite{HM,MN1,MN3,MNN} for its ${\cal N}=4$ generalizations).
Namely, for the ABJM theory with $N_1\le N_2$, the grand canonical partition function is defined by
\begin{align}
\Xi_{k,M}(z)=\sum_{N=0}^\infty z^N Z_k(N,N+M),
\label{gc}
\end{align}
with the partition function $Z_k(N_1,N_2)=Z_k(N,N+M)$ $(M\ge 0)$ and the fugacity $z$.

It is interesting that the corresponding brane configurations on a line enjoy symmetries of the HW transitions \cite{HW}.
Namely, if we indicate NS5-branes and $(1,k)$5-branes by $\bullet$ and $\circ$ respectively and specify the numbers of D3-branes in the intervals, the brane configurations are symmetric under the exchange of 5-branes
\begin{align}
\cdots K\begin{array}{c}\bullet\\[-8pt]\circ\end{array}L\begin{array}{c}\circ\\[-8pt]\bullet\end{array}M\cdots&=\cdots K\begin{array}{c}\circ\\[-8pt]\bullet\end{array}K-L+M+k\begin{array}{c}\bullet\\[-8pt]\circ\end{array}M\cdots,\nonumber\\
\cdots K\begin{array}{c}\bullet\\[-8pt]\circ\end{array}L\begin{array}{c}\bullet\\[-8pt]\circ\end{array}M\cdots&=\cdots K\begin{array}{c}\bullet\\[-8pt]\circ\end{array}K-L+M\begin{array}{c}\bullet\\[-8pt]\circ\end{array}M\cdots,
\label{HW}
\end{align}
with the double symbols $\bullet$ and $\circ$ in the same order.
(See section \ref{propose} for more details on the setup of brane configurations.)
Interestingly, the HW transitions \eqref{HW} can be obtained by requiring conservations of brane charges, such as
\begin{align}
Q_{\text{RR}}=-\frac{k}{2}(\#(1,k)5|_\text{L}-\#(1,k)5|_\text{R})+(\#\text{D3}|_\text{L}-\#\text{D3}|_\text{R}),
\label{charge}
\end{align}
for NS5-branes, where $\#(1,k)5|_\text{L/R}$ is the total number of $(1,k)$5-branes placed on the left/right side while $\#\text{D3}|_\text{L/R}$ is the number of adjacent D3-branes on each side.
It is, however, perplexing if we apply these charge conservations to brane configurations {\it on a circle}, where it does not make sense to distinguish left and right and the interpretation of charge conservations \eqref{charge} is ambiguous.

This observation implies to cut the circle of brane configurations at one specific interval and temporarily forbid the HW transitions there \cite{KM}.
Then, the charge conservations \eqref{charge} are validated on the segment.
Since the idea of cut is similar to fixing a reference frame in mechanics or fixing a local patch in manifolds, we refer to this interval as the reference interval.

Then, we can regard duality cascades to be subsequent applications of the HW transitions in brane configurations on a circle with the reference changed when necessary \cite{FMMN}.
Namely, given an arbitrary brane configuration with D3-branes compactified on a circle and 5-branes located perpendicularly to the D3-branes and tilted relatively to preserve supersymmetries, {\it the working hypothesis} of duality cascades can be formulated as follows.
Namely, let us first assign one of the intervals with the lowest ranks as the reference.
In applying the HW transitions by exchanging 5-branes arbitrarily without crossing the reference, if we encounter lower ranks we relabel references.
We repeat the above step until no lower ranks appear.
(Negative ranks are interpreted as anti-D3-branes.
Since we focus on supersymmetric configurations without negative ranks, or more correctly, the grand canonical partition function \eqref{gc} with sources providing overall ranks, we assume the overall rank $N$ to be large enough.)

Note that, although the HW transitions summarize nicely dualities and duality cascades, the physical dynamics has to be studied separately with caution for each field-theoretical condition, such as dimensions, numbers of supercharges, gauge groups and matter contents.
Indeed, dualities and duality cascades for three-dimensional supersymmetric gauge theories are studied intensively in \cite{ABJ,AHHO,EK,HoKu} and we should identify duality cascades by the above working hypothesis carefully.
Here the dualities implied by the HW transitions are valid with enough numbers of supercharges and we assume that various directions of the HW transitions are all realized in duality cascades.

For the ABJM theory (with two 5-branes), by following duality cascades we continue to flow to lower ranks.
In general for brane configurations with multiple ranks, however, duality cascades can be non-trivial with ambiguities of the directions of ``lower'' ranks and complexities of infinite processes of the HW transitions.
Our above working hypothesis successively separates the infinity (changing references) from finite processes by continuing to assign the lowest ranks to be the reference.
The idea of assigning the lowest ranks to be the reference originates from the studies of grand canonical partition functions of the ABJM theory and its ${\cal N}=4$ generalizations.
For example, in the analysis in the so-called closed string formalism \cite{KMZ,H,MS2,MN5,closed}, the fractional M2-branes are regarded as small fluctuations around the full-fledged M2-branes to be integrated out.
This supports the viewpoint of separating the lowest overall rank from the relative ranks.

Then, for the above working hypothesis of duality cascades, it is natural to ask the following questions.
\begin{itemize}
\item
Starting from an arbitrary brane configuration, does the duality cascade always end?
\item
Does the final destination of duality cascades depend only on the initial brane configuration regardless of processes of duality cascades?
\end{itemize}
Since duality cascades reduce ranks (of finite positive integers), it always ends into either supersymmetric or non-supersymmetric configurations apparently.
However, since we are concentrating on the grand canonical partition function of supersymmetric configurations \eqref{gc} with sources providing overall ranks, the first question becomes non-trivial if we are allowed to add overall ranks freely when we encounter negative ranks.
Also, the second question should be considered if we assume that various directions of duality cascades are all realized.
These are the most fundamental and important physical questions for the above working hypothesis of duality cascades in brane configurations on a circle and have been studied for various different field-theoretical conditions (see for example \cite{SDC}).
Here we intend to add some insights to the questions.

Assuming these questions are answered positively, it is natural to consider the parameter domain of relative ranks consisting of all of brane configurations with no more duality cascades applicable.
Since we expect that all supersymmetric configurations reduce to this domain after applying duality cascades, let us refer to it as the fundamental domain of supersymmetric brane configurations in duality cascades, or simply {\it the fundamental domain} in the following.
Then, it is also interesting to figure out the geometrical properties of the fundamental domain, such as connectedness, compactness and convexity.

Here duality cascades are realized by changing references and, from the arguments of charge conservations \eqref{charge}, they can be interpreted as translations in the parameter space of relative ranks \cite{KM}.
Then, the above questions on the finiteness and the uniqueness of duality cascades can be reformulated geometrically as follows \cite{FMMN}.
\begin{itemize}
\item
Given a point in the parameter space of relative ranks, after applying various processes of the translations corresponding to changing references, does it always reduce to the fundamental domain?
Conversely, starting from the fundamental domain, do all of its copies obtained by the translations cover the whole parameter space?
\item
Do various processes of the translations lead to one point uniquely in the fundamental domain after applying the translations?
Conversely, starting from the fundamental domain, after we apply the translations, are all its distinct copies disjoint, sharing at most the boundaries?
\end{itemize}
Namely, our physical questions whether duality cascades always end and whether the endpoint of duality cascades is unique reduce to the geometrical question whether the fundamental domain can tile the whole parameter space of relative ranks by translations, or in other words, whether the fundamental domain is {\it a parallelotope}.

The brane configurations we have considered have interesting aspects of quantum curves \cite{MiMo,ACDKV}.
Especially, the brane configurations associated to the del Pezzo geometries of genus one enjoy large symmetries of exceptional Weyl groups.
These group-theoretical structures have played central roles in the studies of the matrix models \cite{MNY,KMN,KM,M,FMS,FMN,MY,FMMN}.
For those with interpretations of the del Pezzo geometries enjoying symmetries of Weyl groups, the above questions on duality cascades were answered by discovering the structure of affine Weyl groups \cite{FMMN}.
Namely, the fundamental domain in the Weyl chamber is nothing but the affine Weyl chamber and the translations in duality cascades are those constructed by affine Weyl reflections.

The arguments in \cite{FMMN}, however, apply only to the situations with symmetries of Weyl groups.
For those generally without symmetries of Weyl groups, these questions remain unanswered.
In this paper, we turn to these cases.
Instead of the structure of affine Weyl groups, we utilize theories of combinatorial geometries, such as zonotopes \cite{polytope}.
We shall argue that, given a combination of 5-branes on a circle, we can always associate it to a parallelotope by considering the fundamental domain.
Note that here the interior points of parallelotopes correspond to the relative ranks obtained in duality cascades and the changes of references in duality cascades are realized by the discrete translations for parallelotopes.
It is interesting that the reference introduced originally from charge conservations \eqref{charge} leads to the discrete translations for parallelotopes.

Interestingly, relations of $a$-maximizations \cite{IW} to zonotopes were studied previously in \cite{K} and relations of duality cascades were studied in \cite{EF} for quiver gauge theories obtained from brane tilings \cite{HaKe,FHKVW}.
Comparatively, the setup of our brane configurations consists only of flat branes in the flat spacetime and we expect that the simple structure may be more intuitive and fruitful.
Indeed, it passes through zonotopes and leads directly to parallelotopes.
Besides, note that, for our case, zonotopes serve simply as technical tools to study parallelotopes.
For example, in the brane configuration with symmetries of the $F_4$ Weyl group \cite{FMMN}, the fundamental domain is the 24-cell which is a four-dimensional parallelotope but not a zonotope.
It should be interesting to clarify relations to parallelotopes also for the setup of brane tilings.

This paper is organized as follows.
In the next section, after clarifying the setup of brane configurations and reviewing duality cascades, we state our main proposal.
After studying the proposal for brane configurations with only four 5-branes explicitly in section \ref{four}, we turn to those with arbitrary numbers of 5-branes in section \ref{more}.
Finally, we conclude with summaries and discussions on future directions.

\section{Fundamental domains as parallelotopes}\label{propose}

In this section, after explaining brane configurations and reviewing duality cascades, from our physical motivations, we state our main proposal that the fundamental domain (of supersymmetric brane configurations in duality cascades) is a parallelotope.

\subsection{Brane configurations}

We first explain the setup of brane configurations \cite{HW} considered in this paper (see table \ref{configurations}).
We compactify direction 6 on a circle and consider brane configurations with $(p,q)$5-branes placed perpendicularly at various positions on the circle, where the $(p,q)$5-brane is a bound state of $p$ NS5-branes and $q$ D5-branes.
To preserve supersymmetries, the $(p,q)$5-branes extend to directions of 3,4,5 tilted to directions 7,8,9 respectively by a common angle $\arctan q/p$, besides directions of 0,1,2.
Without breaking supersymmetries, we can insert various numbers of D3-branes which extend in directions 0,1,2,6 and end on pairs of $(p,q)$5-branes.

\begin{table}[t!]
\begin{center}
\begin{tabular}{c|ccccccc}
directions&0&1&2&6&3\;\;\;\;7&4\;\;\;\;8&5\;\;\;\;9\\\hline
D3-branes&$-$&$-$&$-$&$-$&&&\\
NS5-branes&$-$&$-$&$-$&&$-\;\;\;\;\;$&$-\;\;\;\;\;$&$-\;\;\;\;\;$\\
D5-branes&$-$&$-$&$-$&&$\;\;\;\;\;-$&$\;\;\;\;\;-$&$\;\;\;\;\;-$\\
$(1,k)$5-branes&$-$&$-$&$-$&&[3,7]$_k$&[4,8]$_k$&[5,9]$_k$\\
$(p,q)$5-branes&$-$&$-$&$-$&&[3,7]$_{\frac{q}{p}}$&[4,8]$_{\frac{q}{p}}$&[5,9]$_{\frac{q}{p}}$
\end{tabular}
\end{center}
\caption{Directions where various branes in IIB string theory extend.
Here, $[3,7]_k$, for example, denotes direction 3 tilted to direction 7 by an angle $\arctan k$.}
\label{configurations}
\end{table}

\begin{figure}[t!]
\centering\includegraphics[scale=0.6,angle=-90]{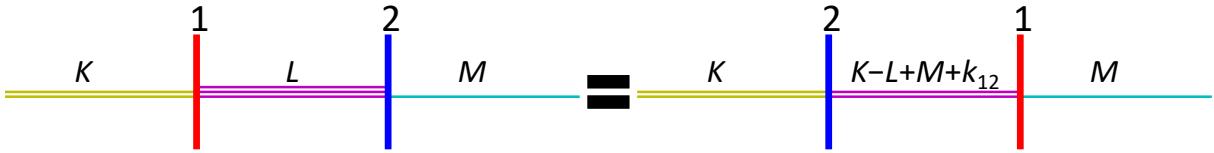}
\caption{A brane cartoon for the HW transitions \eqref{HWpq}.
The horizontal direction denotes direction 6, while the vertical direction denotes directions $3,4,5$ and directions $7,8,9$ totally.}
\label{cartoon}
\end{figure}

The HW transitions \cite{HW,BHKK} indicate that, when we exchange two 5-branes with charges $(p_1,q_1)$ and $(p_2,q_2)$, $k_{12}$ D3-branes are generated with
\begin{align}
k_{12}=\bigg|\det\begin{pmatrix}p_1&p_2\\q_1&q_2\end{pmatrix}\bigg|,
\label{leveldet}
\end{align}
depending on the charges of the two 5-branes.
See figure \ref{cartoon} for a brane cartoon of the HW transitions.
If we denote the two 5-branes by\footnote{The notation used in the introduction is $\bullet=\stackrel{(1,0)}{\bullet}$ and $\circ=\stackrel{(1,k)}{\bullet}$.} $\stackrel{(p_1,q_1)}{\bullet}$ and $\stackrel{(p_2,q_2)}{\bullet}$, the Hanany-Witten transitions exchanging these two 5-branes are expressed by
\begin{align}
\cdots K\stackrel{(p_1,q_1)}{\bullet}L\stackrel{(p_2,q_2)}{\bullet}M\cdots
=\cdots K\stackrel{(p_2,q_2)}{\bullet}K-L+M+k_{12}\stackrel{(p_1,q_1)}{\bullet}M\cdots,
\label{HWpq}
\end{align}
which can be regarded as a shorthand notation for the brane cartoon in figure \ref{cartoon}.
The corresponding field theories are read off by identifying the number of D3-branes in each interval as the rank of unitary groups (with the level $\pm k_{12}$) along with pairs of bifundamental matters connecting adjacent unitary groups \cite{IK}.
The HW transitions originate from studies of dualities in three-dimensional supersymmetric gauge theories \cite{HW}.
Although similar dualities apply to other gauge theories, we need to pay caution to various field-theoretical conditions in applications.

\subsection{Duality cascades}\label{dc}

\begin{figure}[t!]
\centering\includegraphics[scale=0.6,angle=-90]{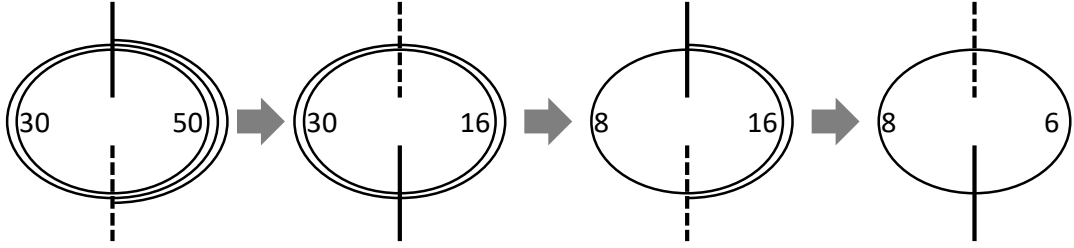}
\caption{A brane cartoon for duality cascades in the ABJM theory.
The circular direction denotes direction 6 while the vertical direction denotes directions 3,4,5 and directions 7,8,9 totally.
The solid vertical line denotes NS5-branes, while the dashed one denotes $(1,k)$5-branes.
In the process of duality cascades, ranks decrease in general.}
\label{abjmdc}
\end{figure}

After reviewing brane configurations and the HW transitions in the previous subsection, let us apply the HW transitions to various brane configurations on a circle.
We first consider the brane configuration in the leftmost figure of figure \ref{abjmdc}, which corresponds to the ABJM theory with gauge group $\text{U}(30)_{6}\times\text{U}(50)_{-6}$ (with the subscripts denoting the levels, $k=6$).
We can apply the HW transitions \eqref{HWpq} to the brane configuration locally by equating $K$ and $M$ since these two variables refer to the numbers of D3-branes in the same stack.
If we apply the HW transitions by setting $K=M=30$ and $L=50$, we find that the number of D3-branes in between changes into $K-L+M+k=16$.
If we further apply by setting $K=M=30$ and $L=16$, the brane configuration returns to the original one.
Instead, we can apply reversely by setting $K=M=16$ and $L=30$ to find that the brane configuration continues to decrease ranks.
By continuing the HW transitions alternatingly, finally we arrive at the brane configuration in the rightmost figure of figure \ref{abjmdc} corresponding to gauge group $\text{U}(6)_{6}\times\text{U}(8)_{-6}$.
In this sense, by subsequent applications of the HW transitions, we are led generally to lower ranks.
If we try to further continue, ranks start to increase.
Note that as discussed in \cite{ABJ} duality cascades to lower ranks indicate the non-unitarity in the field-theoretical description.
Hence, we naturally expect that duality cascades end into the above gauge group with the lowest possible ranks.

\begin{figure}[t!]
\centering\includegraphics[scale=0.6,angle=-90]{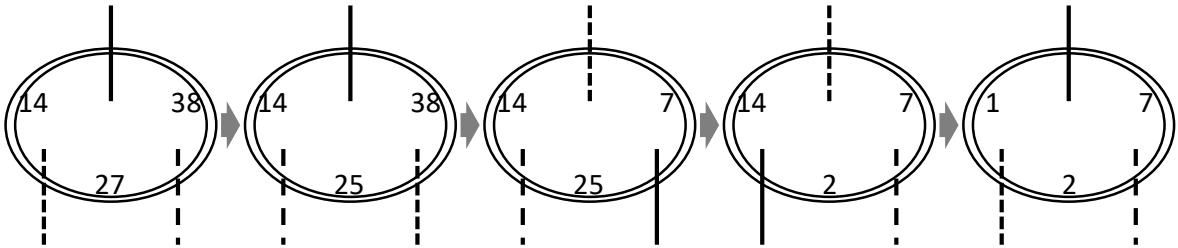}
\caption{A brane cartoon for duality cascades in another brane configuration with one NS5-branes and two $(1,k)$5-branes.
While both of dashed vertical lines denote $(1,k)$5-branes, we distinguish them by different lines.
With multiple relative ranks, the direction of duality cascades to lower ranks is not very clear.}
\label{dcexample}
\end{figure}

The situation is much more ambiguous if we have more 5-branes with more ranks in the brane configuration.
Next, let us consider another brane configuration with one NS5-brane and two $(1,k)$5-branes in the leftmost figure of figure \ref{dcexample} with $k=6$, which corresponds to the three-dimensional ${\cal N}=4$ supersymmetric Chern-Simons theory with gauge group $\text{U}(14)_6\times\text{U}(38)_{-6}\times\text{U}(27)_0$.
The two $(1,k)$5-branes are distinguished by different dashed lines in the figure.
We subsequently apply the HW transitions to the brane configuration.
In the first step we exchange the two $(1,k)$5-branes by setting $K=14$, $L=27$ and $K=38$ to find $K-L+M=25$.
Similarly, we continue several steps of exchanging 5-branes by applying the HW transitions to different pairs of adjacent 5-branes locally.
Finally we arrive at the rightmost figure of figure \ref{dcexample}, whose corresponding gauge group is $\text{U}(1)_6\times\text{U}(7)_{-6}\times\text{U}(2)_0$.

At first sight it is not clear whether we can continue the HW transitions to reduce ranks of the brane configuration.
It is not even clear what we mean by reducing ranks when there are multiple ranks.
Also, we would like to know whether we can reach brane configurations at the end, where duality cascades occur no more.
If we assume that various directions of duality cascades are all realized, we can consider other processes of duality cascades.
Namely, although we have chosen the above process of 5-brane exchanges, apriori there are no reasons to stick to it.
Then, it is also unclear whether we are led to totally different configurations or not by following other processes.
Note that, although the HW transitions may lead to negative ranks, we are allowed to add overall ranks freely since we are considering the grand canonical partition function of supersymmetric brane configurations \eqref{gc} with sources of overall ranks.

Due to these ambiguities, we need to first fix the exact rule of duality cascades.
Let us propose the working hypothesis of duality cascades as follows.
\begin{enumerate}
\item
Indicate one of the intervals with the lowest ranks as the reference.
\item
Apply the HW transitions \eqref{HW} by exchanging 5-branes arbitrarily without crossing the reference.
\item
If we encounter lower ranks, relabel one of the lowest ranks as the reference.
\item
Continue the last two steps until we find no lower ranks in the HW transitions.
\end{enumerate}
Note that this working hypothesis enjoys many properties we expect for duality cascades;
(i) it is consistent with the expectation that ranks (and the degrees of freedom) decrease,
(ii) it covers almost all possible processes of duality cascades (with exceptions of the HW transitions crossing the reference which generally increase ranks),
(iii) it fixes the ambiguities of left and right in charge conservations \eqref{charge} with references and
(iv) it is consistent with the closed string formalism \cite{KMZ,H,MS2,MN5,closed} (where fractional branes are regarded as fluctuations around full branes).
For these reasons, we believe that at present it is the most natural viewpoint of duality cascades.

With the working hypothesis of duality cascades, let us introduce the notation by indicating the reference by the endpoint of the bracket $\langle\cdots\rangle$ and expressing 5-branes by $\bullet=\stackrel{(1,0)}{\bullet}$ and $\stackrel{1}{\circ}{=}\stackrel{2}{\circ}{=}\stackrel{(1,k)}{\bullet}$, where we label the two $(1,k)$5-branes by Arabic numerals.
For example, the brane configuration in the leftmost figure of figure \ref{dcexample} is represented as $\langle 14\stackrel{}{\bullet}38\stackrel{1}{\circ}27\stackrel{2}{\circ}\rangle$.
Then, we can repeat the same process of duality cascades in figure \ref{dcexample} with this notation.
The process of duality cascades in figure \ref{dcexample} is expressed in the first row of table \ref{twocascades}.
Namely, starting with the brane configuration $\langle 14\stackrel{}{\bullet}38\stackrel{1}{\circ}27\stackrel{2}{\circ}\rangle$, by bringing $\stackrel{2}{\circ}$ to the leftmost using the HW transitions, we find $\langle 14\stackrel{2}{\circ}7\stackrel{}{\bullet}25\stackrel{1}{\circ}\rangle$.
Since the lowest rank changes, we change references by rotating cyclically into $\langle 7\stackrel{}{\bullet}25\stackrel{1}{\circ}14\stackrel{2}{\circ}\rangle$.
We can apply the HW transitions again to move to $\langle 7\stackrel{1}{\circ}2\stackrel{2}{\circ}1\stackrel{}{\bullet}\rangle$ and change into $\langle 1\stackrel{}{\bullet}7\stackrel{1}{\circ}2\stackrel{2}{\circ}\rangle$ by cyclic rotations.
Apriori, this is not the unique process.
For example, alternatively in the second row, starting from the same brane configuration, we can bring $\bullet$ to the rightmost to find $\langle 14\stackrel{1}{\circ}9\stackrel{2}{\circ}2\stackrel{}{\bullet}\rangle$ and cyclically rotate into $\langle 2\stackrel{}{\bullet}14\stackrel{1}{\circ}9\stackrel{2}{\circ}\rangle$.
Then, we move to $\langle 2\stackrel{2}{\circ}1\stackrel{}{\bullet}7\stackrel{1}{\circ}\rangle$, which surprisingly reduces to the same final configuration by rotations.
It is unclear in general whether the final brane configuration depends only on the initial one regardless of processes of duality cascades.
(Of course there are many other processes available which are covered by the working hypothesis and should also be considered.)

\begin{table}[t!]
\begin{center}
\begin{tabular}{c}
$\rotatebox[origin=c]{30}{=}\;
\langle 14\stackrel{2}{\circ}7\stackrel{}{\bullet}25\stackrel{1}{\circ}\rangle
\to\langle 7\stackrel{}{\bullet}25\stackrel{1}{\circ}14\stackrel{2}{\circ}\rangle
=\langle 7\stackrel{1}{\circ}2\stackrel{2}{\circ}1\stackrel{}{\bullet}\rangle
\;\rotatebox[origin=c]{-30}{$\to$}$\\
$\langle 14\stackrel{}{\bullet}38\stackrel{1}{\circ}27\stackrel{2}{\circ}\rangle\hspace{75mm}
\langle 1\stackrel{}{\bullet}7\stackrel{1}{\circ}2\stackrel{2}{\circ}\rangle$\\
$\rotatebox[origin=c]{-30}{=}\;
\langle 14\stackrel{1}{\circ}9\stackrel{2}{\circ}2\stackrel{}{\bullet}\rangle
\to\langle 2\stackrel{}{\bullet}14\stackrel{1}{\circ}9\stackrel{2}{\circ}\rangle
=\langle 2\stackrel{2}{\circ}1\stackrel{}{\bullet}7\stackrel{1}{\circ}\rangle
\;\rotatebox[origin=c]{30}{$\to$}$
\end{tabular}
\end{center}
\caption{Two processes of duality cascades.
Here we indicate NS5-branes and $(1,k)$5-branes by $\bullet$ and $\circ$ respectively (with labels of Arabic numerals), specify the number of D3-branes in each interval and denote the reference by the endpoint of the bracket.
Also, $=$ denotes the HW transitions with the reference fixed, while $\to$ denotes the cyclic rotations changing references.
Both processes reduce exactly to the same configuration.}
\label{twocascades}
\end{table}

To summarize for now, besides the brane configuration of the ABJM theory with only one relative rank, generally for those with multiple ranks, the concept of duality cascades is fixed \cite{FMMN} by the working hypothesis of duality cascades, which enjoys many expected properties for duality cascades.
After fixing the working hypothesis, it is natural to ask the following questions.
\begin{itemize}
\item
Do the processes of duality cascades always end?
\item
Does the final brane configuration depend only on the initial one regardless of processes?
\end{itemize}
Although we choose a specific combination of 5-branes to explain the question, this should be answered for arbitrary combinations of 5-branes.

\subsection{Fundamental domains}\label{conj}

\begin{figure}[t!]
\centering\includegraphics[scale=0.6,angle=-90]{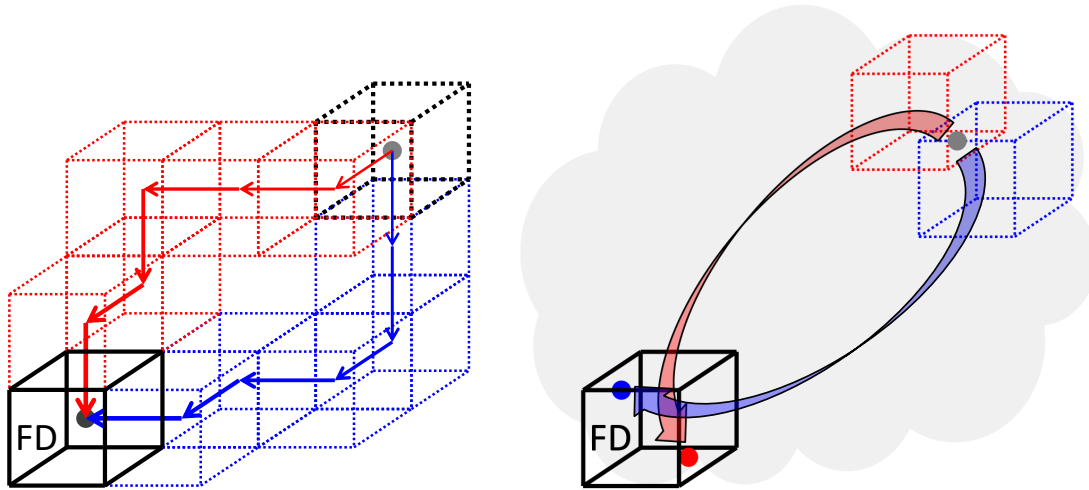}
\caption{Relations between duality cascades and parallelotopes.
With several translations for changing references provided, duality cascades are realized as subsequent translations.
(Left)
If we define the fundamental domain (FD) by collecting final destinations of duality cascades, the question whether duality cascades always end is translated into whether FD covers the whole space by translations and the question whether the endpoint of duality cascades is unique is translated into whether all the distinct copies of FD are disjoint.
Especially, various processes lead to the unique endpoint only if distinct copies do not overlap, since the position in the copies of FD is invariant under processes of duality cascades.
(Right)
If the red box and the blue box (obtained from two different processes of translations for FD) overlap, the same point in the overlap can lead to both the red point and the blue point in FD from two different processes of translations.}
\label{fdparallelotope}
\end{figure}

In the previous subsection, we have provided examples of duality cascades.
For the working hypothesis of duality cascades, it is natural to ask whether duality cascades always end and whether the final brane configuration depends only on the initial one regardless of processes of duality cascades.
As explained in \cite{FMMN} and repeated in the introduction, to formulate the questions more explicitly, we first collect all the brane configurations where duality cascades do not occur any more and refer to it as the fundamental domain (of supersymmetric brane configurations in duality cascades).
Then, since changing references in duality cascades is realized by translations in the parameter space of relative ranks \cite{KM}, after providing the translations corresponding to duality cascades, the questions can be restated geometrically in terms of the fundamental domain as follows (see figure \ref{fdparallelotope}).

\begin{itemize}
\item
Starting with a point in the parameter space of relative ranks, after applying the translations (corresponding to duality cascades), does it always reduce to the fundamental domain?
Conversely, do all the copies of the fundamental domain obtained by the translations cover the whole parameter space of relative ranks?
\item
Do different processes of the translations lead to one point uniquely in the fundamental domain?
Conversely, are all the distinct copies of the fundamental domain disjoint sharing at most their boundaries?
\end{itemize}
In short, the question is whether the fundamental domain forms a parallelotope.

As noted in the previous subsection, physically it was discussed \cite{ABJ} for the ABJM theory that the failure of locating in the fundamental domain indicates the non-unitarity in the field-theoretical description.
It is then natural to expect that, by following duality cascades, we continue to flow to better descriptions and finally arrive at unitary theories uniquely.
For this reason, it is natural to expect that duality cascades always end and the endpoint is unique regardless of processes of duality cascades.
In other words, by the above arguments, we conjecture the fundamental domains to be parallelotopes.
Note that the relative ranks can be parametrized arbitrarily, which are related by affine transformations.
Whether a polytope is a parallelotope, however, does not depend on the parameterizations.

In the following, we provide our arguments for the claim that the fundamental domains are parallelotopes for arbitrary numbers of 5-branes.
To avoid our general arguments from being too abstract at the first sight, we first provide a concrete computation with four 5-branes in section \ref{four}, which is also instructive for later generalizations in section \ref{more}.

\section{Four 5-branes}\label{four}

To argue the validity of our proposal that the fundamental domain is a parallelotope, let us consider brane configurations with only four 5-branes in this section, starting with the case with all 5-branes different.
After introducing alternative descriptions for the fundamental domain known as zonotopes, we can argue the proposal of parallelotopes with nice mathematical properties of zonotopes.
We also clarify the physical interpretations for our arguments.
Then, we can turn to cases with some of 5-branes identical by taking degenerate limits.

\subsection{Fundamental domain}

Let us first repeat the proposal that the fundamental domain is a parallelotope explicitly in our current setup with four 5-branes, where the fundamental domain is defined by collecting all the brane configurations where duality cascades do not occur any more.

\begin{table}[t!]
\begin{center}
\begin{tabular}{c|c}
splits&inequalities\\\hline
$\langle\stackrel{\text{1}}{\bullet}\cdot\stackrel{\text{2}}{\bullet}\stackrel{\text{3}}{\bullet}\stackrel{\text{4}}{\bullet}\rangle$&
$M_1-M_2-M_3\le(k_{12}+k_{13}+k_{14})/2$\\
$\langle\stackrel{\text{2}}{\bullet}\cdot\stackrel{\text{1}}{\bullet}\stackrel{\text{3}}{\bullet}\stackrel{\text{4}}{\bullet}\rangle$&
$-M_1+M_2-M_3\le(k_{12}+k_{23}+k_{24})/2$\\
$\langle\stackrel{\text{3}}{\bullet}\cdot\stackrel{\text{1}}{\bullet}\stackrel{\text{2}}{\bullet}\stackrel{\text{4}}{\bullet}\rangle$&
$-M_1-M_2+M_3\le(k_{13}+k_{23}+k_{34})/2$\\
$\langle\stackrel{\text{4}}{\bullet}\cdot\stackrel{\text{1}}{\bullet}\stackrel{\text{2}}{\bullet}\stackrel{\text{3}}{\bullet}\rangle$&
$M_1+M_2+M_3\le(k_{14}+k_{24}+k_{34})/2$\\
$\langle\stackrel{\text{1}}{\bullet}\stackrel{\text{2}}{\bullet}\cdot\stackrel{\text{3}}{\bullet}\stackrel{\text{4}}{\bullet}\rangle$&
$-2M_3\le(k_{13}+k_{23}+k_{14}+k_{24})/2$\\
$\langle\stackrel{\text{1}}{\bullet}\stackrel{\text{3}}{\bullet}\cdot\stackrel{\text{2}}{\bullet}\stackrel{\text{4}}{\bullet}\rangle$&
$-2M_2\le(k_{12}+k_{23}+k_{14}+k_{34})/2$\\
$\langle\stackrel{\text{1}}{\bullet}\stackrel{\text{4}}{\bullet}\cdot\stackrel{\text{2}}{\bullet}\stackrel{\text{3}}{\bullet}\rangle$&
$2M_1\le(k_{12}+k_{13}+k_{24}+k_{34})/2$\\
$\langle\stackrel{\text{2}}{\bullet}\stackrel{\text{3}}{\bullet}\cdot\stackrel{\text{1}}{\bullet}\stackrel{\text{4}}{\bullet}\rangle$&
$-2M_1\le(k_{12}+k_{13}+k_{24}+k_{34})/2$\\
$\langle\stackrel{\text{2}}{\bullet}\stackrel{\text{4}}{\bullet}\cdot\stackrel{\text{1}}{\bullet}\stackrel{\text{3}}{\bullet}\rangle$&
$2M_2\le(k_{12}+k_{23}+k_{14}+k_{34})/2$\\
$\langle\stackrel{\text{3}}{\bullet}\stackrel{\text{4}}{\bullet}\cdot\stackrel{\text{1}}{\bullet}\stackrel{\text{2}}{\bullet}\rangle$&
$2M_3\le(k_{13}+k_{23}+k_{14}+k_{24})/2$\\
$\langle\stackrel{\text{1}}{\bullet}\stackrel{\text{2}}{\bullet}\stackrel{\text{3}}{\bullet}\cdot\stackrel{\text{4}}{\bullet}\rangle$&
$-M_1-M_2-M_3\le(k_{14}+k_{24}+k_{34})/2$\\
$\langle\stackrel{\text{1}}{\bullet}\stackrel{\text{2}}{\bullet}\stackrel{\text{4}}{\bullet}\cdot\stackrel{\text{3}}{\bullet}\rangle$&
$M_1+M_2-M_3\le(k_{13}+k_{23}+k_{34})/2$\\
$\langle\stackrel{\text{1}}{\bullet}\stackrel{\text{3}}{\bullet}\stackrel{\text{4}}{\bullet}\cdot\stackrel{\text{2}}{\bullet}\rangle$&
$M_1-M_2+M_3\le(k_{12}+k_{23}+k_{24})/2$\\
$\langle\stackrel{\text{2}}{\bullet}\stackrel{\text{3}}{\bullet}\stackrel{\text{4}}{\bullet}\cdot\stackrel{\text{1}}{\bullet}\rangle$&
$-M_1+M_2+M_3\le(k_{12}+k_{13}+k_{14})/2$
\end{tabular}
\end{center}
\caption{Inequalities obtained from the HW transitions.
Each inequality is obtained by comparing the reference rank with a specific rank denoted by $\cdot$ with four 5-branes split differently.}
\label{cuboctainequalities}
\end{table}

We first consider brane configurations with all 5-branes different.
Namely, we introduce $(p_i,q_i)$5-branes $(i=1,\cdots,4)$ abbreviated as $\stackrel{i}{\bullet}(=\stackrel{(p_i,q_i)}{\bullet})$ and parameterize the ranks by
\begin{align}
&\langle N\stackrel{\text{1}}{\bullet}
N-M_1+M_2+M_3+(k_{12}+k_{13}+k_{14})/2\stackrel{\text{2}}{\bullet}\nonumber\\
&\quad N+2M_3+(k_{13}+k_{23}+k_{14}+k_{24})/2\stackrel{\text{3}}{\bullet}
N+M_1+M_2+M_3+(k_{14}+k_{24}+k_{34})/2\stackrel{\text{4}}{\bullet}\rangle,
\label{3M}
\end{align}
(in the notation introduced in section \ref{dc}) where $N$ is the overall rank and $(M_1,M_2,M_3)$ are the relative ranks.
This brane configuration describes the three-dimensional ${\cal N}=3$ supersymmetric Chern-Simons theory with gauge group
\begin{align}
&\text{U}(N)_{-k_{14}}\times\text{U}(N-M_1+M_2+M_3+(k_{12}+k_{13}+k_{14})/2)_{k_{12}}\nonumber\\
&\times\text{U}(N+2M_3+(k_{13}+k_{23}+k_{14}+k_{24})/2)_{k_{23}}\nonumber\\
&\times\text{U}(N+M_1+M_2+M_3+(k_{14}+k_{24}+k_{34})/2)_{k_{34}},
\end{align}
and four pairs of bifundamental matters connecting adjacent unitary groups cyclically.
The levels in the subscripts are subject to sign ambiguities due to the definition with absolute values in \eqref{leveldet}.
Here, though the ranks (and the numbers of D3-branes) are subject to the ambiguity of half integers, we shift the parameters by $k_{ij}$ so that the center of the fundamental domain is located at the origin, as discussed below.

We can exchange the order of 5-branes in \eqref{3M} arbitrarily without crossing the reference using the HW transitions \eqref{HWpq}.
From the definition of the fundamental domain, by requiring the reference rank $N$ continues to be the lowest rank in all the HW transitions, we find 14 inequalities totally
\begin{align}
&\pm 2M_1\le(k_{12}+k_{13}+k_{24}+k_{34})/2,\quad
\pm(-M_1+M_2+M_3)\le(k_{12}+k_{13}+k_{14})/2,\nonumber\\
&\pm 2M_2\le(k_{12}+k_{23}+k_{14}+k_{34})/2,\quad
\pm(M_1-M_2+M_3)\le(k_{12}+k_{23}+k_{24})/2,\nonumber\\
&\pm 2M_3\le(k_{13}+k_{23}+k_{14}+k_{24})/2,\quad
\pm(M_1+M_2-M_3)\le(k_{13}+k_{23}+k_{34})/2,\nonumber\\
&\pm(M_1+M_2+M_3)\le(k_{14}+k_{24}+k_{34})/2.
\label{cuboctaineq}
\end{align}

\begin{figure}[t!]
\centering\includegraphics[scale=0.7,angle=-90]{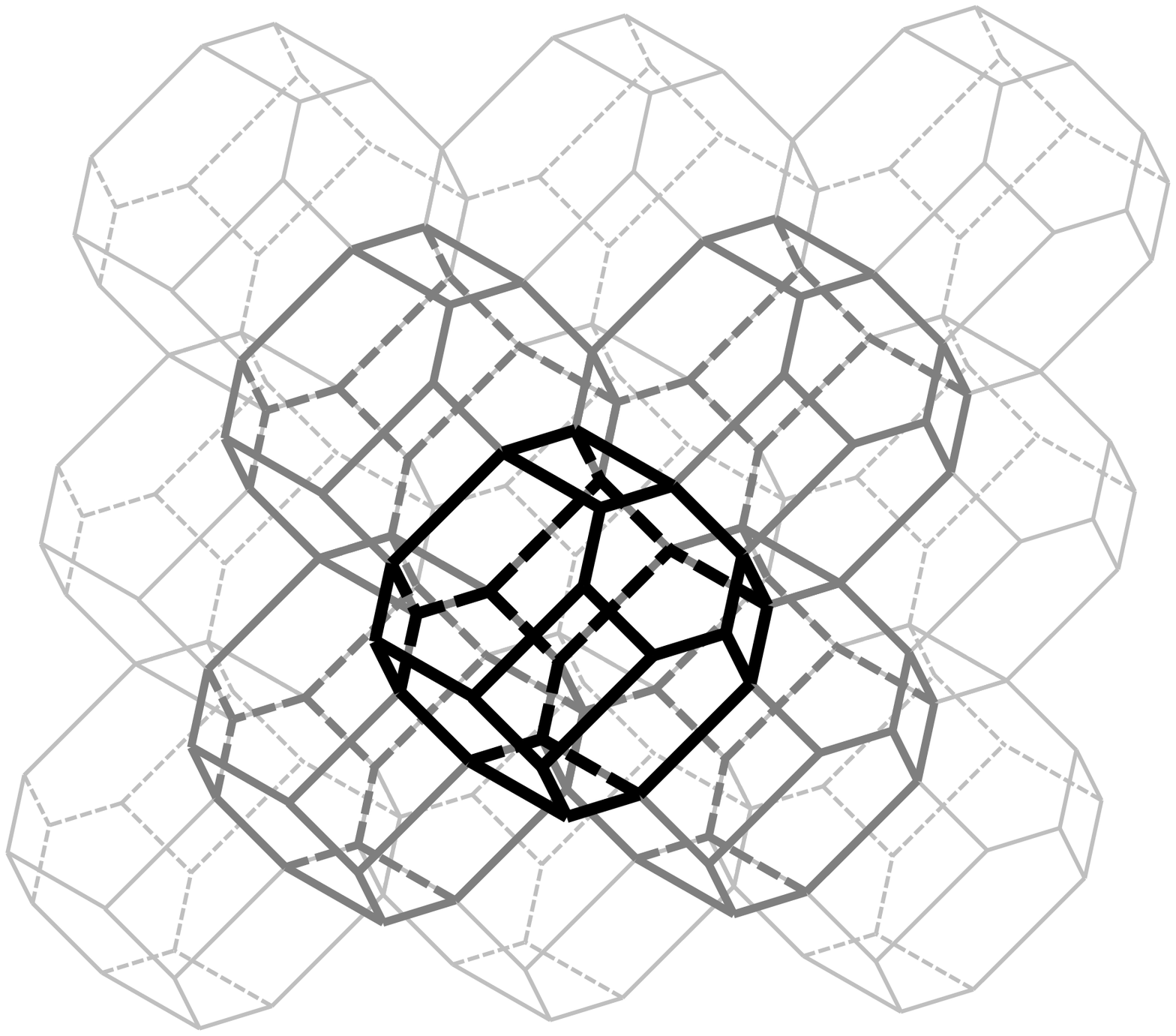}
\caption{The fundamental domain defined by the inequalities \eqref{cuboctaineq} in the parameter space of relative ranks ${\bm M}=(M_1,M_2,M_3)\in{\mathbb R}^3$.
It is combinatorially equivalent to the truncated octahedron and can tile the entire space ${\mathbb R}^3$ by translations.}
\label{tilingfigure}
\end{figure}

Note that each inequality is obtained by comparing the reference rank with the rank of an interval in an order.
After fixing an interval the orders of 5-branes before and after the interval do not matter.
Hence, the inequalities are labeled by how we split the four 5-branes into two disjoint groups, such as $\{1,2,3,4\}=\{1,3\}\sqcup\{2,4\}$.
In other words, if we indicate by $\cdot$ the interval to be compared with the reference, with a slight generalization of the notation in \eqref{3M}, we can label the inequalities by $\langle\stackrel{\text{1}}{\bullet}\stackrel{\text{3}}{\bullet}\cdot\stackrel{\text{2}}{\bullet}\stackrel{\text{4}}{\bullet}\rangle$ for the above example with $\stackrel{\text{1}}{\bullet}$/$\stackrel{\text{3}}{\bullet}$ (and $\stackrel{\text{2}}{\bullet}$/$\stackrel{\text{4}}{\bullet}$ respectively) interchangeable.
This is also understood from the interpretation of charge conservations \eqref{charge}.
By removing the trivial ones with all or none 5-branes in between ($\langle\stackrel{\text{1}}{\bullet}\stackrel{\text{2}}{\bullet}\stackrel{\text{3}}{\bullet}\stackrel{\text{4}}{\bullet}\cdot\rangle$ and $\langle\cdot\stackrel{\text{1}}{\bullet}\stackrel{\text{2}}{\bullet}\stackrel{\text{3}}{\bullet}\stackrel{\text{4}}{\bullet}\rangle$), the number of the inequalities is $2^4-2=14$.
We list the correspondence between the splits and the inequalities in table \ref{cuboctainequalities}.
For the above example of $\langle\stackrel{\text{1}}{\bullet}\stackrel{\text{3}}{\bullet}\cdot\stackrel{\text{2}}{\bullet}\stackrel{\text{4}}{\bullet}\rangle$, we apply the HW transitions and find $\langle N\stackrel{\text{1}}{\bullet}\cdots\stackrel{\text{3}}{\bullet}N+2M_2+(k_{12}+k_{23}+k_{14}+k_{34})/2\stackrel{\text{2}}{\bullet}\cdots\stackrel{\text{4}}{\bullet}\rangle$ with irrelevant ranks omitted. 
By comparing the reference with the rank in the interval denoted by $\cdot$ in $\langle\stackrel{\text{1}}{\bullet}\stackrel{\text{3}}{\bullet}\cdot\stackrel{\text{2}}{\bullet}\stackrel{\text{4}}{\bullet}\rangle$, we find the inequality $N\le N+2M_2+(k_{12}+k_{23}+k_{14}+k_{34})/2$.

Since each inequality restricts to a half space, all of them define a polytope.
This description of the fundamental domain obtained directly from duality cascades leads later to the ${\cal H}$ description in section \ref{equiv}.
Our conjecture stated in section \ref{conj} is that this polytope is a parallelotope.
Namely, we expect that the infinite copies of the polytope obtained by translations can tile the whole parameter space of relative ranks ${\bm M}=(M_1,M_2,M_3)\in{\mathbb R}^3$, which seems to be the case (see figure \ref{tilingfigure}).
In the following, we clarify the structure for our current setup of brane configurations with four 5-branes.
These clarifications are helpful in our generalizations to those with arbitrary numbers of 5-branes in the next section.

\subsection{Descriptions}\label{descriptions}

\begin{table}[t!]
\begin{center}
\begin{tabular}{l|c}
configurations&$(\epsilon_{x},\epsilon_{y},\epsilon_{z},\zeta_{x},\zeta_{y},\zeta_{z})$\\\hline
$\langle\stackrel{\text{1}}{\bullet}\stackrel{\text{2}}{\bullet}\stackrel{\text{3}}{\bullet}\stackrel{\text{4}}{\bullet}\rangle$&
$(-,+,-,-,-,-)$\\
$\langle\stackrel{\text{1}}{\bullet}\stackrel{\text{2}}{\bullet}\stackrel{\text{4}}{\bullet}\stackrel{\text{3}}{\bullet}\rangle=\langle\stackrel{\text{1}}{\bullet}\stackrel{\text{2}}{\bullet}\stackrel{\text{3}}{\bullet}k_{34}\stackrel{\text{4}}{\bullet}\rangle$&
$(-,+,-,-,-,+)$\\
$\langle\stackrel{\text{1}}{\bullet}\stackrel{\text{3}}{\bullet}\stackrel{\text{2}}{\bullet}\stackrel{\text{4}}{\bullet}\rangle=\langle\stackrel{\text{1}}{\bullet}\stackrel{\text{2}}{\bullet}k_{23}\stackrel{\text{3}}{\bullet}\stackrel{\text{4}}{\bullet}\rangle$&
$(+,+,-,-,-,-)$\\
$\langle\stackrel{\text{1}}{\bullet}\stackrel{\text{3}}{\bullet}\stackrel{\text{4}}{\bullet}\stackrel{\text{2}}{\bullet}\rangle=\langle\stackrel{\text{1}}{\bullet}\stackrel{\text{2}}{\bullet}k_{23}+k_{24}\stackrel{\text{3}}{\bullet}k_{24}\stackrel{\text{4}}{\bullet}\rangle$&
$(+,+,-,-,+,-)$\\
$\langle\stackrel{\text{1}}{\bullet}\stackrel{\text{4}}{\bullet}\stackrel{\text{2}}{\bullet}\stackrel{\text{3}}{\bullet}\rangle=\langle\stackrel{\text{1}}{\bullet}\stackrel{\text{2}}{\bullet}k_{24}\stackrel{\text{3}}{\bullet}k_{24}+k_{34}\stackrel{\text{4}}{\bullet}\rangle$&
$(-,+,-,-,+,+)$\\
$\langle\stackrel{\text{1}}{\bullet}\stackrel{\text{4}}{\bullet}\stackrel{\text{3}}{\bullet}\stackrel{\text{2}}{\bullet}\rangle=\langle\stackrel{\text{1}}{\bullet}\stackrel{\text{2}}{\bullet}k_{23}+k_{24}\stackrel{\text{3}}{\bullet}k_{24}+k_{34}\stackrel{\text{4}}{\bullet}\rangle$&
$(+,+,-,-,+,+)$\\
$\langle\stackrel{\text{2}}{\bullet}\stackrel{\text{1}}{\bullet}\stackrel{\text{3}}{\bullet}\stackrel{\text{4}}{\bullet}\rangle=\langle\stackrel{\text{1}}{\bullet}k_{12}\stackrel{\text{2}}{\bullet}\stackrel{\text{3}}{\bullet}\stackrel{\text{4}}{\bullet}\rangle$&
$(-,+,+,-,-,-)$\\
$\langle\stackrel{\text{2}}{\bullet}\stackrel{\text{1}}{\bullet}\stackrel{\text{4}}{\bullet}\stackrel{\text{3}}{\bullet}\rangle=\langle\stackrel{\text{1}}{\bullet}k_{12}\stackrel{\text{2}}{\bullet}\stackrel{\text{3}}{\bullet}k_{34}\stackrel{\text{4}}{\bullet}\rangle$&
$(-,+,+,-,-,+)$\\
$\langle\stackrel{\text{2}}{\bullet}\stackrel{\text{3}}{\bullet}\stackrel{\text{1}}{\bullet}\stackrel{\text{4}}{\bullet}\rangle=\langle\stackrel{\text{1}}{\bullet}k_{12}+k_{13}\stackrel{\text{2}}{\bullet}k_{13}\stackrel{\text{3}}{\bullet}\stackrel{\text{4}}{\bullet}\rangle$&
$(-,-,+,-,-,-)$\\
$\langle\stackrel{\text{2}}{\bullet}\stackrel{\text{3}}{\bullet}\stackrel{\text{4}}{\bullet}\stackrel{\text{1}}{\bullet}\rangle=\langle\stackrel{\text{1}}{\bullet}k_{12}+k_{13}+k_{14}\stackrel{\text{2}}{\bullet}k_{13}+k_{14}\stackrel{\text{3}}{\bullet}k_{14}\stackrel{\text{4}}{\bullet}\rangle$&
$(-,-,+,+,-,-)$\\
$\langle\stackrel{\text{2}}{\bullet}\stackrel{\text{4}}{\bullet}\stackrel{\text{1}}{\bullet}\stackrel{\text{3}}{\bullet}\rangle=\langle\stackrel{\text{1}}{\bullet}k_{12}+k_{14}\stackrel{\text{2}}{\bullet}k_{14}\stackrel{\text{3}}{\bullet}k_{14}+k_{34}\stackrel{\text{4}}{\bullet}\rangle$&
$(-,+,+,+,-,+)$\\
$\langle\stackrel{\text{2}}{\bullet}\stackrel{\text{4}}{\bullet}\stackrel{\text{3}}{\bullet}\stackrel{\text{1}}{\bullet}\rangle=\langle\stackrel{\text{1}}{\bullet}k_{12}+k_{13}+k_{14}\stackrel{\text{2}}{\bullet}k_{13}+k_{14}\stackrel{\text{3}}{\bullet}k_{14}+k_{34}\stackrel{\text{4}}{\bullet}\rangle$&
$(-,-,+,+,-,+)$\\
$\langle\stackrel{\text{3}}{\bullet}\stackrel{\text{1}}{\bullet}\stackrel{\text{2}}{\bullet}\stackrel{\text{4}}{\bullet}\rangle=\langle\stackrel{\text{1}}{\bullet}k_{13}\stackrel{\text{2}}{\bullet}k_{13}+k_{23}\stackrel{\text{3}}{\bullet}\stackrel{\text{4}}{\bullet}\rangle$&
$(+,-,-,-,-,-)$\\
$\langle\stackrel{\text{3}}{\bullet}\stackrel{\text{1}}{\bullet}\stackrel{\text{4}}{\bullet}\stackrel{\text{2}}{\bullet}\rangle=\langle\stackrel{\text{1}}{\bullet}k_{13}\stackrel{\text{2}}{\bullet}k_{13}+k_{23}+k_{24}\stackrel{\text{3}}{\bullet}k_{24}\stackrel{\text{4}}{\bullet}\rangle$&
$(+,-,-,-,+,-)$\\
$\langle\stackrel{\text{3}}{\bullet}\stackrel{\text{2}}{\bullet}\stackrel{\text{1}}{\bullet}\stackrel{\text{4}}{\bullet}\rangle=\langle\stackrel{\text{1}}{\bullet}k_{12}+k_{13}\stackrel{\text{2}}{\bullet}k_{13}+k_{23}\stackrel{\text{3}}{\bullet}\stackrel{\text{4}}{\bullet}\rangle$&
$(+,-,+,-,-,-)$\\
$\langle\stackrel{\text{3}}{\bullet}\stackrel{\text{2}}{\bullet}\stackrel{\text{4}}{\bullet}\stackrel{\text{1}}{\bullet}\rangle=\langle\stackrel{\text{1}}{\bullet}k_{12}+k_{13}+k_{14}\stackrel{\text{2}}{\bullet}k_{13}+k_{23}+k_{14}\stackrel{\text{3}}{\bullet}k_{14}\stackrel{\text{4}}{\bullet}\rangle$&
$(+,-,+,+,-,-)$\\
$\langle\stackrel{\text{3}}{\bullet}\stackrel{\text{4}}{\bullet}\stackrel{\text{1}}{\bullet}\stackrel{\text{2}}{\bullet}\rangle=\langle\stackrel{\text{1}}{\bullet}k_{13}+k_{14}\stackrel{\text{2}}{\bullet}k_{13}+k_{23}+k_{14}+k_{24}\stackrel{\text{3}}{\bullet}k_{14}+k_{24}\stackrel{\text{4}}{\bullet}\rangle$&
$(+,-,-,+,+,-)$\\
$\langle\stackrel{\text{3}}{\bullet}\stackrel{\text{4}}{\bullet}\stackrel{\text{2}}{\bullet}\stackrel{\text{1}}{\bullet}\rangle=\langle\stackrel{\text{1}}{\bullet}k_{12}+k_{13}+k_{14}\stackrel{\text{2}}{\bullet}k_{13}+k_{23}+k_{14}+k_{24}\stackrel{\text{3}}{\bullet}k_{14}+k_{24}\stackrel{\text{4}}{\bullet}\rangle$&
$(+,-,+,+,+,-)$\\
$\langle\stackrel{\text{4}}{\bullet}\stackrel{\text{1}}{\bullet}\stackrel{\text{2}}{\bullet}\stackrel{\text{3}}{\bullet}\rangle=\langle\stackrel{\text{1}}{\bullet}k_{14}\stackrel{\text{2}}{\bullet}k_{14}+k_{24}\stackrel{\text{3}}{\bullet}k_{14}+k_{24}+k_{34}\stackrel{\text{4}}{\bullet}\rangle$&
$(-,+,-,+,+,+)$\\
$\langle\stackrel{\text{4}}{\bullet}\stackrel{\text{1}}{\bullet}\stackrel{\text{3}}{\bullet}\stackrel{\text{2}}{\bullet}\rangle=\langle\stackrel{\text{1}}{\bullet}k_{14}\stackrel{\text{2}}{\bullet}k_{23}+k_{14}+k_{24}\stackrel{\text{3}}{\bullet}k_{14}+k_{24}+k_{34}\stackrel{\text{4}}{\bullet}\rangle$&
$(+,+,-,+,+,+)$\\
$\langle\stackrel{\text{4}}{\bullet}\stackrel{\text{2}}{\bullet}\stackrel{\text{1}}{\bullet}\stackrel{\text{3}}{\bullet}\rangle=\langle\stackrel{\text{1}}{\bullet}k_{12}+k_{14}\stackrel{\text{2}}{\bullet}k_{14}+k_{24}\stackrel{\text{3}}{\bullet}k_{14}+k_{24}+k_{34}\stackrel{\text{4}}{\bullet}\rangle$&
$(-,+,+,+,+,+)$\\
$\langle\stackrel{\text{4}}{\bullet}\stackrel{\text{2}}{\bullet}\stackrel{\text{3}}{\bullet}\stackrel{\text{1}}{\bullet}\rangle=\langle\stackrel{\text{1}}{\bullet}k_{12}+k_{13}+k_{14}\stackrel{\text{2}}{\bullet}k_{13}+k_{14}+k_{24}\stackrel{\text{3}}{\bullet}k_{14}+k_{24}+k_{34}\stackrel{\text{4}}{\bullet}\rangle$&
$(-,-,+,+,+,+)$\\
$\langle\stackrel{\text{4}}{\bullet}\stackrel{\text{3}}{\bullet}\stackrel{\text{1}}{\bullet}\stackrel{\text{2}}{\bullet}\rangle=\langle\stackrel{\text{1}}{\bullet}k_{13}+k_{14}\stackrel{\text{2}}{\bullet}k_{13}+k_{23}+k_{14}+k_{24}\stackrel{\text{3}}{\bullet}k_{14}+k_{24}+k_{34}\stackrel{\text{4}}{\bullet}\rangle$&
$(+,-,-,+,+,+)$\\
$\langle\stackrel{\text{4}}{\bullet}\stackrel{\text{3}}{\bullet}\stackrel{\text{2}}{\bullet}\stackrel{\text{1}}{\bullet}\rangle=\langle\stackrel{\text{1}}{\bullet}k_{12}+k_{13}+k_{14}\stackrel{\text{2}}{\bullet}k_{13}+k_{23}+k_{14}+k_{24}\stackrel{\text{3}}{\bullet}k_{14}+k_{24}+k_{34}\stackrel{\text{4}}{\bullet}\rangle$&
$(+,-,+,+,+,+)$
\end{tabular}
\end{center}
\caption{Vertices of the fundamental domain, which are branes configurations without rank differences.
We bring 5-branes in brane configurations without rank differences into the standard order by the HW transitions (with vanishing relative ranks 0 omitted).
The resulting relative ranks ${\bm M}$ in \eqref{3M} can be expressed as \eqref{Msigns} by specifying the signs $(\epsilon_x,\epsilon_y,\epsilon_z,\zeta_x,\zeta_y,\zeta_z)$.}
\label{cuboctasigns}
\end{table}

It is not simple, however, to proceed with these inequalities.
We shall provide a few alternative descriptions for the fundamental domain in the following (which lead to more general constructions in section \ref{equiv}).
Instead of the inequalities, it was argued \cite{KM,FMMN} that we may understand this polytope from its vertices (leading to the ${\cal V}$ description) by introducing brane configurations without rank differences but in various orders of 5-branes.
It is not clear at all at the first sight that the convex hull of them is equivalent to the fundamental domain defined by the inequalities \eqref{cuboctaineq}.
This is the case, however, from explicit constructions here or more general arguments given later in section \ref{equiv}.

Let us bring the brane configurations without rank differences into the standard order of \eqref{3M} using the HW transitions and read off the relative ranks ${\bm M}=(M_1,M_2,M_3)$ by comparison.
For example for the brane configuration $\langle N\stackrel{\text{3}}{\bullet}N\stackrel{\text{1}}{\bullet}N\stackrel{\text{4}}{\bullet}N\stackrel{\text{2}}{\bullet}\rangle$ with the overall rank $N$ omitted as $\langle\stackrel{\text{3}}{\bullet}\stackrel{\text{1}}{\bullet}\stackrel{\text{4}}{\bullet}\stackrel{\text{2}}{\bullet}\rangle$, we bring the 5-branes into the standard order of \eqref{3M} using the HW transitions \eqref{HWpq} and find 
$\langle\stackrel{\text{1}}{\bullet}k_{13}\stackrel{\text{2}}{\bullet}k_{13}+k_{23}+k_{24}\stackrel{\text{3}}{\bullet}k_{24}\stackrel{\text{4}}{\bullet}\rangle$.
By comparing this with the parameterization in \eqref{3M}, we can read off the parameters for relative ranks ${\bm M}=(M_1,M_2,M_3)$.

Note that, in table \ref{cuboctasigns}, in bringing the 5-branes into the standard order, negative ranks of $k_{ij}$, or lower ranks compared with the reference, never appear.
Since both the original order and the standard order can be arbitrary, the brane configurations without rank differences are always located within the fundamental domain defined by the inequalities \eqref{cuboctaineq}.
This fact is important in relating generally the polytope defined by inequalities (in the ${\cal H}$ description) and the polytope defined as the convex hull of brane configurations without rank differences (in the ${\cal V}$ description).

It is then interesting to find that, by introducing the vectors
\begin{align}
&{\bm u}_{x}=\frac{k_{23}}{4}(0,-1,1),\quad
{\bm u}_{y}=\frac{k_{13}}{4}(1,0,-1),\quad
{\bm u}_{z}=\frac{k_{12}}{4}(-1,1,0),\nonumber\\
&{\bm v}_{x}=\frac{k_{14}}{4}(0,1,1),\quad
{\bm v}_{y}=\frac{k_{24}}{4}(1,0,1),\quad
{\bm v}_{z}=\frac{k_{34}}{4}(1,1,0),
\label{uv}
\end{align}
all the relative ranks ${\bm M}$ for the brane configurations without rank differences are given by
\begin{align}
{\bm V}=\epsilon_x{\bm u}_{x}+\epsilon_y{\bm u}_{y}+\epsilon_z{\bm u}_{z}
+\zeta_{x}{\bm v}_{x}+\zeta_{y}{\bm v}_{y}+\zeta_{z}{\bm v}_{z},
\label{Msigns}
\end{align}
with $(\epsilon_x,\epsilon_y,\epsilon_z,\zeta_x,\zeta_y,\zeta_z)$ denoting signs $\pm 1$.
For the above example, $\langle\stackrel{\text{3}}{\bullet}\stackrel{\text{1}}{\bullet}\stackrel{\text{4}}{\bullet}\stackrel{\text{2}}{\bullet}\rangle=\langle\stackrel{\text{1}}{\bullet}k_{13}\stackrel{\text{2}}{\bullet}k_{13}+k_{23}+k_{24}\stackrel{\text{3}}{\bullet}k_{24}\stackrel{\text{4}}{\bullet}\rangle$, we can express the resulting relative ranks ${\bm M}$ with the vectors \eqref{uv} by ${\bm V}=+{\bm u}_{x}-{\bm u}_{y}-{\bm u}_{z}-{\bm v}_{x}+{\bm v}_{y}-{\bm v}_{z}$, from which we can then read off signs $(\epsilon_x,\epsilon_y,\epsilon_z,\zeta_x,\zeta_y,\zeta_z)$ in \eqref{Msigns}.
In table \ref{cuboctasigns} we list all signs we read off from the analysis.

Besides, it is interesting to find that, in bringing the 5-branes into the standard order, the parameters $k_{ij}$ appear in a clear pattern.
Namely, by comparing the original order of 5-branes and the standard order, if the order of two 5-branes $i$ and $j$ are reversed, $k_{ij}$ connecting the 5-branes $i$ and $j$ appears in the relative ranks.
For example, if we compare the above example $\langle\stackrel{\text{3}}{\bullet}\stackrel{\text{1}}{\bullet}\stackrel{\text{4}}{\bullet}\stackrel{\text{2}}{\bullet}\rangle$ with the standard order $\langle\stackrel{\text{1}}{\bullet}\stackrel{\text{2}}{\bullet}\stackrel{\text{3}}{\bullet}\stackrel{\text{4}}{\bullet}\rangle$, we find that the orders between the pairs of $\stackrel{\text{1}}{\bullet}$/$\stackrel{\text{3}}{\bullet}$, $\stackrel{\text{2}}{\bullet}$/$\stackrel{\text{3}}{\bullet}$ and  $\stackrel{\text{2}}{\bullet}$/$\stackrel{\text{4}}{\bullet}$ are reversed.
For this reason, in the final expression $\langle\stackrel{\text{1}}{\bullet}k_{13}\stackrel{\text{2}}{\bullet}k_{13}+k_{23}+k_{24}\stackrel{\text{3}}{\bullet}k_{24}\stackrel{\text{4}}{\bullet}\rangle$, $k_{13}$, $k_{23}$ and $k_{24}$ appear connecting the corresponding 5-branes.
In other words, we may want to summarize this property schematically as
\begin{align}
&[\stackrel{\text{3}}{\bullet},\stackrel{\text{2}}{\bullet}]
=(\stackrel{\text{1}}{\bullet}\stackrel{\text{2}}{\bullet}k_{23}\stackrel{\text{3}}{\bullet}\stackrel{\text{4}}{\bullet})=2{\bm u}_x,\quad
[\stackrel{\text{4}}{\bullet},\stackrel{\text{1}}{\bullet}]
=(\stackrel{\text{1}}{\bullet}k_{14}\stackrel{\text{2}}{\bullet}k_{14}\stackrel{\text{3}}{\bullet}k_{14}\stackrel{\text{4}}{\bullet})=2{\bm v}_x,
\nonumber\\
&[\stackrel{\text{3}}{\bullet},\stackrel{\text{1}}{\bullet}]
=(\stackrel{\text{1}}{\bullet}k_{13}\stackrel{\text{2}}{\bullet}k_{13}\stackrel{\text{3}}{\bullet}\stackrel{\text{4}}{\bullet})=-2{\bm u}_y,\quad
[\stackrel{\text{4}}{\bullet},\stackrel{\text{2}}{\bullet}]
=(\stackrel{\text{1}}{\bullet}\stackrel{\text{2}}{\bullet}k_{24}\stackrel{\text{3}}{\bullet}k_{24}\stackrel{\text{4}}{\bullet})=2{\bm v}_y,\nonumber\\
&[\stackrel{\text{2}}{\bullet},\stackrel{\text{1}}{\bullet}]
=(\stackrel{\text{1}}{\bullet}k_{12}\stackrel{\text{2}}{\bullet}\stackrel{\text{3}}{\bullet}\stackrel{\text{4}}{\bullet})=2{\bm u}_z,\quad
[\stackrel{\text{4}}{\bullet},\stackrel{\text{3}}{\bullet}]
=(\stackrel{\text{1}}{\bullet}\stackrel{\text{2}}{\bullet}\stackrel{\text{3}}{\bullet}k_{34}\stackrel{\text{4}}{\bullet})=2{\bm v}_z.
\label{commutation}
\end{align}
Here the exchanges of 5-branes and the resulting shifts of relative ranks are expressed respectively in the commutation relations on the left and the vectors on the right, while in the middle we introduce an intermediate expression of the shifting vectors in the basis related directly to brane configurations
\begin{align}
{\bm M}=(\stackrel{\text{1}}{\bullet}-M_1+M_2+M_3\stackrel{\text{2}}{\bullet}2M_3\stackrel{\text{3}}{\bullet}M_1+M_2+M_3\stackrel{\text{4}}{\bullet}).
\end{align}

\begin{figure}[t!]
\centering\includegraphics[scale=0.75,angle=-90]{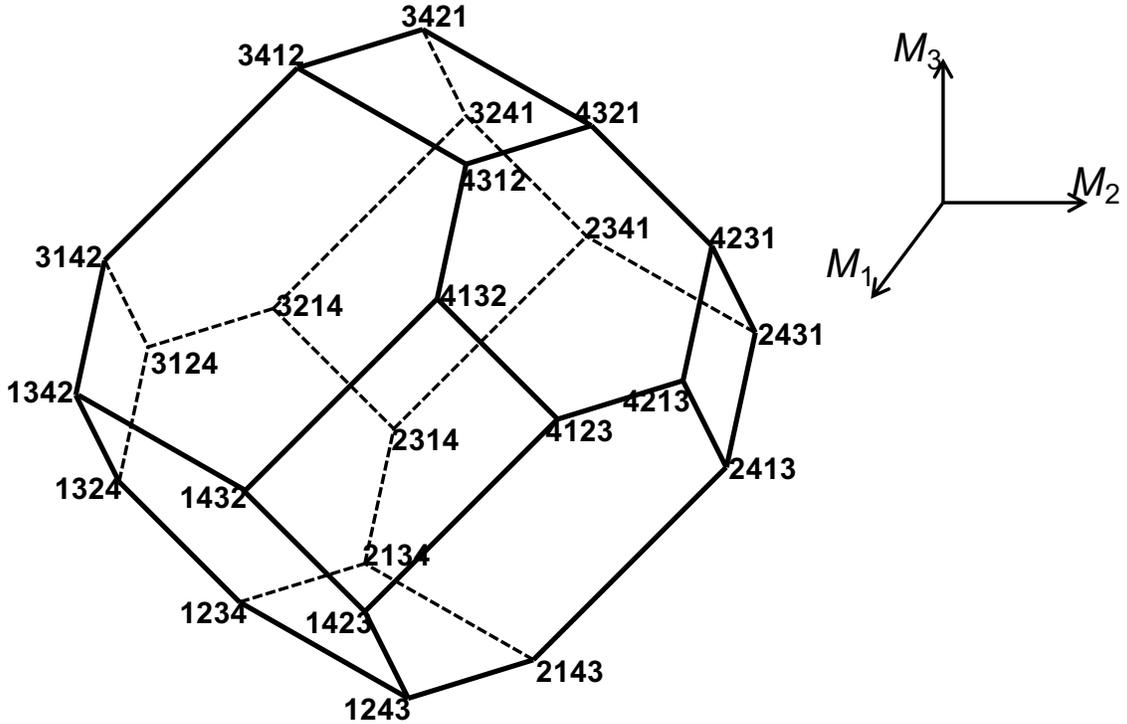}
\caption{The polytope of the fundamental domain.
Instead of the original description by the inequalities \eqref{cuboctaineq}, we can characterize the polytope from its vertices which are labeled by the order of 5-branes.}
\label{cuboctafigure}
\end{figure}

We plot the convex hull in figure \ref{cuboctafigure} with the vertices labelled by the orders of 5-branes.
The polytope is combinatorially equivalent to the truncated octahedron and known to tile the space of ${\bm M}=(M_1,M_2,M_3)\in{\mathbb R}^3$ as in figure \ref{tilingfigure}.
Interestingly, exactly the same figure appears, for example, in figure 15.1.5 in \cite{handbook}.
This implies relations to a zonotope (besides, to a permutohedron, as noted later) in discrete geometries \cite{polytope}, which is the Minkowski sum of vectors called generators.
Namely, with the terminology of zonotopes, the fundamental domain \eqref{cuboctaineq} is generated by the vectors $(2{\bm u}_x,2{\bm u}_y,2{\bm u}_z,2{\bm v}_x,2{\bm v}_y,2{\bm v}_z)$.
This leads to the ${\cal Z}$ description in section \ref{equiv}.
Note that each edge corresponds to the vector exchanging two 5-branes in \eqref{commutation}.
This is because the brane configurations exchanging only two 5-branes share enough numbers of inequalities.

Also, note that vertices of each face (labeled by splitting 5-branes into two groups, as inequalities) are direct products of all permutations of 5-branes in the two groups.
For example, the face $\langle\stackrel{\text{1}}{\bullet}\stackrel{\text{4}}{\bullet}\cdot\stackrel{\text{2}}{\bullet}\stackrel{\text{3}}{\bullet}\rangle$ is a parallelogram having 4 vertices constructed by direct products of permutations $\{\stackrel{\text{1}}{\bullet}\stackrel{\text{4}}{\bullet},\stackrel{\text{4}}{\bullet}\stackrel{\text{1}}{\bullet}\}$ and $\{\stackrel{\text{2}}{\bullet}\stackrel{\text{3}}{\bullet},\stackrel{\text{3}}{\bullet}\stackrel{\text{2}}{\bullet}\}$, while the face $\langle\stackrel{\text{4}}{\bullet}\cdot\stackrel{\text{1}}{\bullet}\stackrel{\text{2}}{\bullet}\stackrel{\text{3}}{\bullet}\rangle$ is a hexagon having 6 vertices which are trivial direct products of $\{\stackrel{\text{4}}{\bullet}\}$ and 6 permutations $\{\stackrel{\text{1}}{\bullet}\stackrel{\text{2}}{\bullet}\stackrel{\text{3}}{\bullet},\cdots\}$ (see figure \ref{cuboctafigure}).
The splitting is because the inequality is satisfied marginally on the face.

Here for brane configurations with only four 5-branes, we have introduced a few equivalent descriptions explicitly.
In section \ref{equiv}, to study brane configurations with arbitrary numbers of 5-branes, we shall generalize them into the three descriptions (${\cal H}$, ${\cal Z}$, ${\cal V}$) and argue their equivalence generally.
Various facts found here will be helpful.

\subsection{Parallelotope}\label{signmattest}

Using mainly the description by zonotopes, let us study whether the fundamental domain is a parallelotope in this subsection.
It is a famous classical result for zonotopes \cite{Shephard,McMullen} that this question can be tested by whether the rank of the sign matrix for centers of faces coincides with the dimension.

Since all faces of zonotopes are centrally symmetric, the center of each face is obtained by averaging its vertices.
From the results of table \ref{cuboctasigns}, it is not difficult to find that the centers are also of the form
\begin{align}
{\bm F}=\bar\epsilon_x{\bm u}_{x}+\bar\epsilon_y{\bm u}_{y}+\bar\epsilon_z{\bm u}_{z}
+\bar\zeta_{x}{\bm v}_{x}+\bar\zeta_{y}{\bm v}_{y}+\bar\zeta_{z}{\bm v}_{z},
\label{centersign}
\end{align}
but now some components in $(\bar\epsilon_x,\bar\epsilon_y,\bar\epsilon_z,\bar\zeta_x,\bar\zeta_y,\bar\zeta_z)$ can be vanishing.
For example, since the four vertices on the face $\langle\stackrel{\text{1}}{\bullet}\stackrel{\text{3}}{\bullet}\cdot\stackrel{\text{2}}{\bullet}\stackrel{\text{4}}{\bullet}\rangle$ are given by $+{\bm u}_{x}\pm{\bm u}_{y}-{\bm u}_{z}-{\bm v}_{x}\pm{\bm v}_{y}-{\bm v}_{z}$ (with any double sign), their average is $+{\bm u}_{x}-{\bm u}_{z}-{\bm v}_{x}-{\bm v}_{z}$.
We list the average of the vertices on each face in table \ref{cuboctafaces}.

\begin{table}[t!]
\begin{center}
\begin{tabular}{c|c|c}
faces&vertices&$(\bar\epsilon_x,\bar\epsilon_y,\bar\epsilon_z,\bar\zeta_x,\bar\zeta_y,\bar\zeta_z)$\\\hline
$\langle\stackrel{\text{1}}{\bullet}\cdot\stackrel{\text{2}}{\bullet}\stackrel{\text{3}}{\bullet}\stackrel{\text{4}}{\bullet}\rangle$&
$\langle\stackrel{\text{1}}{\bullet}\stackrel{\text{2}}{\bullet}\stackrel{\text{3}}{\bullet}\stackrel{\text{4}}{\bullet}\rangle$,
$\langle\stackrel{\text{1}}{\bullet}\stackrel{\text{2}}{\bullet}\stackrel{\text{4}}{\bullet}\stackrel{\text{3}}{\bullet}\rangle$,
$\langle\stackrel{\text{1}}{\bullet}\stackrel{\text{3}}{\bullet}\stackrel{\text{2}}{\bullet}\stackrel{\text{4}}{\bullet}\rangle$,
$\langle\stackrel{\text{1}}{\bullet}\stackrel{\text{3}}{\bullet}\stackrel{\text{4}}{\bullet}\stackrel{\text{2}}{\bullet}\rangle$,
$\langle\stackrel{\text{1}}{\bullet}\stackrel{\text{4}}{\bullet}\stackrel{\text{2}}{\bullet}\stackrel{\text{3}}{\bullet}\rangle$,
$\langle\stackrel{\text{1}}{\bullet}\stackrel{\text{4}}{\bullet}\stackrel{\text{3}}{\bullet}\stackrel{\text{2}}{\bullet}\rangle$&
$(\;0,+,-,-,\;0,\;0)$\\
$\langle\stackrel{\text{2}}{\bullet}\cdot\stackrel{\text{1}}{\bullet}\stackrel{\text{3}}{\bullet}\stackrel{\text{4}}{\bullet}\rangle$&
$\langle\stackrel{\text{2}}{\bullet}\stackrel{\text{1}}{\bullet}\stackrel{\text{3}}{\bullet}\stackrel{\text{4}}{\bullet}\rangle$,
$\langle\stackrel{\text{2}}{\bullet}\stackrel{\text{1}}{\bullet}\stackrel{\text{4}}{\bullet}\stackrel{\text{3}}{\bullet}\rangle$,
$\langle\stackrel{\text{2}}{\bullet}\stackrel{\text{3}}{\bullet}\stackrel{\text{1}}{\bullet}\stackrel{\text{4}}{\bullet}\rangle$,
$\langle\stackrel{\text{2}}{\bullet}\stackrel{\text{3}}{\bullet}\stackrel{\text{4}}{\bullet}\stackrel{\text{1}}{\bullet}\rangle$,
$\langle\stackrel{\text{2}}{\bullet}\stackrel{\text{4}}{\bullet}\stackrel{\text{1}}{\bullet}\stackrel{\text{3}}{\bullet}\rangle$,
$\langle\stackrel{\text{2}}{\bullet}\stackrel{\text{4}}{\bullet}\stackrel{\text{3}}{\bullet}\stackrel{\text{1}}{\bullet}\rangle$&
$(-,\;0,+,\;0,-,\;0)$\\
$\langle\stackrel{\text{3}}{\bullet}\cdot\stackrel{\text{1}}{\bullet}\stackrel{\text{2}}{\bullet}\stackrel{\text{4}}{\bullet}\rangle$&
$\langle\stackrel{\text{3}}{\bullet}\stackrel{\text{1}}{\bullet}\stackrel{\text{2}}{\bullet}\stackrel{\text{4}}{\bullet}\rangle$,
$\langle\stackrel{\text{3}}{\bullet}\stackrel{\text{1}}{\bullet}\stackrel{\text{4}}{\bullet}\stackrel{\text{2}}{\bullet}\rangle$,
$\langle\stackrel{\text{3}}{\bullet}\stackrel{\text{2}}{\bullet}\stackrel{\text{1}}{\bullet}\stackrel{\text{4}}{\bullet}\rangle$,
$\langle\stackrel{\text{3}}{\bullet}\stackrel{\text{2}}{\bullet}\stackrel{\text{4}}{\bullet}\stackrel{\text{1}}{\bullet}\rangle$,
$\langle\stackrel{\text{3}}{\bullet}\stackrel{\text{4}}{\bullet}\stackrel{\text{1}}{\bullet}\stackrel{\text{2}}{\bullet}\rangle$,
$\langle\stackrel{\text{3}}{\bullet}\stackrel{\text{4}}{\bullet}\stackrel{\text{2}}{\bullet}\stackrel{\text{1}}{\bullet}\rangle$&
$(+,-,\;0,\;0,\;0,-)$\\
$\langle\stackrel{\text{4}}{\bullet}\cdot\stackrel{\text{1}}{\bullet}\stackrel{\text{2}}{\bullet}\stackrel{\text{3}}{\bullet}\rangle$&
$\langle\stackrel{\text{4}}{\bullet}\stackrel{\text{1}}{\bullet}\stackrel{\text{2}}{\bullet}\stackrel{\text{3}}{\bullet}\rangle$,
$\langle\stackrel{\text{4}}{\bullet}\stackrel{\text{1}}{\bullet}\stackrel{\text{3}}{\bullet}\stackrel{\text{2}}{\bullet}\rangle$,
$\langle\stackrel{\text{4}}{\bullet}\stackrel{\text{2}}{\bullet}\stackrel{\text{1}}{\bullet}\stackrel{\text{3}}{\bullet}\rangle$,
$\langle\stackrel{\text{4}}{\bullet}\stackrel{\text{2}}{\bullet}\stackrel{\text{3}}{\bullet}\stackrel{\text{1}}{\bullet}\rangle$,
$\langle\stackrel{\text{4}}{\bullet}\stackrel{\text{3}}{\bullet}\stackrel{\text{1}}{\bullet}\stackrel{\text{2}}{\bullet}\rangle$,
$\langle\stackrel{\text{4}}{\bullet}\stackrel{\text{3}}{\bullet}\stackrel{\text{2}}{\bullet}\stackrel{\text{1}}{\bullet}\rangle$&
$(\;0,\;0,\;0,+,+,+)$\\
$\langle\stackrel{\text{1}}{\bullet}\stackrel{\text{2}}{\bullet}\cdot\stackrel{\text{3}}{\bullet}\stackrel{\text{4}}{\bullet}\rangle$&
$\langle\stackrel{\text{1}}{\bullet}\stackrel{\text{2}}{\bullet}\stackrel{\text{3}}{\bullet}\stackrel{\text{4}}{\bullet}\rangle$,
$\langle\stackrel{\text{1}}{\bullet}\stackrel{\text{2}}{\bullet}\stackrel{\text{4}}{\bullet}\stackrel{\text{3}}{\bullet}\rangle$,
$\langle\stackrel{\text{2}}{\bullet}\stackrel{\text{1}}{\bullet}\stackrel{\text{3}}{\bullet}\stackrel{\text{4}}{\bullet}\rangle$,
$\langle\stackrel{\text{2}}{\bullet}\stackrel{\text{1}}{\bullet}\stackrel{\text{4}}{\bullet}\stackrel{\text{3}}{\bullet}\rangle$&
$(-,+,\;0,-,-,\;0)$\\
$\langle\stackrel{\text{1}}{\bullet}\stackrel{\text{3}}{\bullet}\cdot\stackrel{\text{2}}{\bullet}\stackrel{\text{4}}{\bullet}\rangle$&
$\langle\stackrel{\text{1}}{\bullet}\stackrel{\text{3}}{\bullet}\stackrel{\text{2}}{\bullet}\stackrel{\text{4}}{\bullet}\rangle$,
$\langle\stackrel{\text{1}}{\bullet}\stackrel{\text{3}}{\bullet}\stackrel{\text{4}}{\bullet}\stackrel{\text{2}}{\bullet}\rangle$,
$\langle\stackrel{\text{3}}{\bullet}\stackrel{\text{1}}{\bullet}\stackrel{\text{2}}{\bullet}\stackrel{\text{4}}{\bullet}\rangle$,
$\langle\stackrel{\text{3}}{\bullet}\stackrel{\text{1}}{\bullet}\stackrel{\text{4}}{\bullet}\stackrel{\text{2}}{\bullet}\rangle$&
$(+,\;0,-,-,\;0,-)$\\
$\langle\stackrel{\text{1}}{\bullet}\stackrel{\text{4}}{\bullet}\cdot\stackrel{\text{2}}{\bullet}\stackrel{\text{3}}{\bullet}\rangle$&
$\langle\stackrel{\text{1}}{\bullet}\stackrel{\text{4}}{\bullet}\stackrel{\text{2}}{\bullet}\stackrel{\text{3}}{\bullet}\rangle$,
$\langle\stackrel{\text{1}}{\bullet}\stackrel{\text{4}}{\bullet}\stackrel{\text{3}}{\bullet}\stackrel{\text{2}}{\bullet}\rangle$,
$\langle\stackrel{\text{4}}{\bullet}\stackrel{\text{1}}{\bullet}\stackrel{\text{2}}{\bullet}\stackrel{\text{3}}{\bullet}\rangle$,
$\langle\stackrel{\text{4}}{\bullet}\stackrel{\text{1}}{\bullet}\stackrel{\text{3}}{\bullet}\stackrel{\text{2}}{\bullet}\rangle$&
$(\;0,+,-,\;0,+,+)$\\
$\langle\stackrel{\text{2}}{\bullet}\stackrel{\text{3}}{\bullet}\cdot\stackrel{\text{1}}{\bullet}\stackrel{\text{4}}{\bullet}\rangle$&
$\langle\stackrel{\text{2}}{\bullet}\stackrel{\text{3}}{\bullet}\stackrel{\text{1}}{\bullet}\stackrel{\text{4}}{\bullet}\rangle$,
$\langle\stackrel{\text{2}}{\bullet}\stackrel{\text{3}}{\bullet}\stackrel{\text{4}}{\bullet}\stackrel{\text{1}}{\bullet}\rangle$,
$\langle\stackrel{\text{3}}{\bullet}\stackrel{\text{2}}{\bullet}\stackrel{\text{1}}{\bullet}\stackrel{\text{4}}{\bullet}\rangle$,
$\langle\stackrel{\text{3}}{\bullet}\stackrel{\text{2}}{\bullet}\stackrel{\text{4}}{\bullet}\stackrel{\text{1}}{\bullet}\rangle$&
$(\;0,-,+,\;0,-,-)$\\
$\langle\stackrel{\text{2}}{\bullet}\stackrel{\text{4}}{\bullet}\cdot\stackrel{\text{1}}{\bullet}\stackrel{\text{3}}{\bullet}\rangle$&
$\langle\stackrel{\text{2}}{\bullet}\stackrel{\text{4}}{\bullet}\stackrel{\text{1}}{\bullet}\stackrel{\text{3}}{\bullet}\rangle$,
$\langle\stackrel{\text{2}}{\bullet}\stackrel{\text{4}}{\bullet}\stackrel{\text{3}}{\bullet}\stackrel{\text{1}}{\bullet}\rangle$,
$\langle\stackrel{\text{4}}{\bullet}\stackrel{\text{2}}{\bullet}\stackrel{\text{1}}{\bullet}\stackrel{\text{3}}{\bullet}\rangle$,
$\langle\stackrel{\text{4}}{\bullet}\stackrel{\text{2}}{\bullet}\stackrel{\text{3}}{\bullet}\stackrel{\text{1}}{\bullet}\rangle$&
$(-,\;0,+,+,\;0,+)$\\
$\langle\stackrel{\text{3}}{\bullet}\stackrel{\text{4}}{\bullet}\cdot\stackrel{\text{1}}{\bullet}\stackrel{\text{2}}{\bullet}\rangle$&
$\langle\stackrel{\text{3}}{\bullet}\stackrel{\text{4}}{\bullet}\stackrel{\text{1}}{\bullet}\stackrel{\text{2}}{\bullet}\rangle$,
$\langle\stackrel{\text{3}}{\bullet}\stackrel{\text{4}}{\bullet}\stackrel{\text{2}}{\bullet}\stackrel{\text{1}}{\bullet}\rangle$,
$\langle\stackrel{\text{4}}{\bullet}\stackrel{\text{3}}{\bullet}\stackrel{\text{1}}{\bullet}\stackrel{\text{2}}{\bullet}\rangle$,
$\langle\stackrel{\text{4}}{\bullet}\stackrel{\text{3}}{\bullet}\stackrel{\text{2}}{\bullet}\stackrel{\text{1}}{\bullet}\rangle$&
$(+,-,\;0,+,+,\;0)$\\
$\langle\stackrel{\text{1}}{\bullet}\stackrel{\text{2}}{\bullet}\stackrel{\text{3}}{\bullet}\cdot\stackrel{\text{4}}{\bullet}\rangle$&
$\langle\stackrel{\text{1}}{\bullet}\stackrel{\text{2}}{\bullet}\stackrel{\text{3}}{\bullet}\stackrel{\text{4}}{\bullet}\rangle$,
$\langle\stackrel{\text{1}}{\bullet}\stackrel{\text{3}}{\bullet}\stackrel{\text{2}}{\bullet}\stackrel{\text{4}}{\bullet}\rangle$,
$\langle\stackrel{\text{2}}{\bullet}\stackrel{\text{1}}{\bullet}\stackrel{\text{3}}{\bullet}\stackrel{\text{4}}{\bullet}\rangle$,
$\langle\stackrel{\text{2}}{\bullet}\stackrel{\text{3}}{\bullet}\stackrel{\text{1}}{\bullet}\stackrel{\text{4}}{\bullet}\rangle$,
$\langle\stackrel{\text{3}}{\bullet}\stackrel{\text{1}}{\bullet}\stackrel{\text{2}}{\bullet}\stackrel{\text{4}}{\bullet}\rangle$,
$\langle\stackrel{\text{3}}{\bullet}\stackrel{\text{2}}{\bullet}\stackrel{\text{1}}{\bullet}\stackrel{\text{4}}{\bullet}\rangle$&
$(\;0,\;0,\;0,-,-,-)$\\
$\langle\stackrel{\text{1}}{\bullet}\stackrel{\text{2}}{\bullet}\stackrel{\text{4}}{\bullet}\cdot\stackrel{\text{3}}{\bullet}\rangle$&
$\langle\stackrel{\text{1}}{\bullet}\stackrel{\text{2}}{\bullet}\stackrel{\text{4}}{\bullet}\stackrel{\text{3}}{\bullet}\rangle$,
$\langle\stackrel{\text{1}}{\bullet}\stackrel{\text{4}}{\bullet}\stackrel{\text{2}}{\bullet}\stackrel{\text{3}}{\bullet}\rangle$,
$\langle\stackrel{\text{2}}{\bullet}\stackrel{\text{1}}{\bullet}\stackrel{\text{4}}{\bullet}\stackrel{\text{3}}{\bullet}\rangle$,
$\langle\stackrel{\text{2}}{\bullet}\stackrel{\text{4}}{\bullet}\stackrel{\text{1}}{\bullet}\stackrel{\text{3}}{\bullet}\rangle$,
$\langle\stackrel{\text{4}}{\bullet}\stackrel{\text{1}}{\bullet}\stackrel{\text{2}}{\bullet}\stackrel{\text{3}}{\bullet}\rangle$,
$\langle\stackrel{\text{4}}{\bullet}\stackrel{\text{2}}{\bullet}\stackrel{\text{1}}{\bullet}\stackrel{\text{3}}{\bullet}\rangle$&
$(-,+,\;0,\;0,\;0,+)$\\
$\langle\stackrel{\text{1}}{\bullet}\stackrel{\text{3}}{\bullet}\stackrel{\text{4}}{\bullet}\cdot\stackrel{\text{2}}{\bullet}\rangle$&
$\langle\stackrel{\text{1}}{\bullet}\stackrel{\text{3}}{\bullet}\stackrel{\text{4}}{\bullet}\stackrel{\text{2}}{\bullet}\rangle$,
$\langle\stackrel{\text{1}}{\bullet}\stackrel{\text{4}}{\bullet}\stackrel{\text{3}}{\bullet}\stackrel{\text{2}}{\bullet}\rangle$,
$\langle\stackrel{\text{3}}{\bullet}\stackrel{\text{1}}{\bullet}\stackrel{\text{4}}{\bullet}\stackrel{\text{2}}{\bullet}\rangle$,
$\langle\stackrel{\text{3}}{\bullet}\stackrel{\text{4}}{\bullet}\stackrel{\text{1}}{\bullet}\stackrel{\text{2}}{\bullet}\rangle$,
$\langle\stackrel{\text{4}}{\bullet}\stackrel{\text{1}}{\bullet}\stackrel{\text{3}}{\bullet}\stackrel{\text{2}}{\bullet}\rangle$,
$\langle\stackrel{\text{4}}{\bullet}\stackrel{\text{3}}{\bullet}\stackrel{\text{1}}{\bullet}\stackrel{\text{2}}{\bullet}\rangle$&
$(+,\;0,-,\;0,+,\;0)$\\
$\langle\stackrel{\text{2}}{\bullet}\stackrel{\text{3}}{\bullet}\stackrel{\text{4}}{\bullet}\cdot\stackrel{\text{1}}{\bullet}\rangle$&
$\langle\stackrel{\text{2}}{\bullet}\stackrel{\text{3}}{\bullet}\stackrel{\text{4}}{\bullet}\stackrel{\text{1}}{\bullet}\rangle$,
$\langle\stackrel{\text{2}}{\bullet}\stackrel{\text{4}}{\bullet}\stackrel{\text{3}}{\bullet}\stackrel{\text{1}}{\bullet}\rangle$,
$\langle\stackrel{\text{3}}{\bullet}\stackrel{\text{2}}{\bullet}\stackrel{\text{4}}{\bullet}\stackrel{\text{1}}{\bullet}\rangle$,
$\langle\stackrel{\text{3}}{\bullet}\stackrel{\text{4}}{\bullet}\stackrel{\text{2}}{\bullet}\stackrel{\text{1}}{\bullet}\rangle$,
$\langle\stackrel{\text{4}}{\bullet}\stackrel{\text{2}}{\bullet}\stackrel{\text{3}}{\bullet}\stackrel{\text{1}}{\bullet}\rangle$,
$\langle\stackrel{\text{4}}{\bullet}\stackrel{\text{3}}{\bullet}\stackrel{\text{2}}{\bullet}\stackrel{\text{1}}{\bullet}\rangle$&
$(\;0,-,+,+,\;0,\;0)$
\end{tabular}
\end{center}
\caption{Centers of faces.
The sign for the center of each face $(\bar\epsilon_x,\bar\epsilon_y,\bar\epsilon_z,\bar\zeta_x,\bar\zeta_y,\bar\zeta_z)$ is obtained by averaging the signs for its vertices $(\epsilon_x,\epsilon_y,\epsilon_z,\zeta_x,\zeta_y,\zeta_z)$ given in table \ref{cuboctasigns}.}
\label{cuboctafaces}
\end{table}

Then, according to \cite{Shephard,McMullen}, whether the zonotope is a parallelotope depends on whether the rank of the matrix $E$ listing signs \eqref{centersign} for all the centers matches the space dimension.
For our case the matrix is
\begin{align}
E=\begin{pmatrix}
0&0&0&1&1&1\\
0&-1&1&1&0&0\\
1&0&-1&0&1&0\\
-1&1&0&0&0&1\\
0&1&-1&0&1&1\\
-1&0&1&1&0&1\\
1&-1&0&1&1&0
\end{pmatrix},
\label{fourE}
\end{align}
where we list signs $(\bar\epsilon_x,\bar\epsilon_y,\bar\epsilon_z,\bar\zeta_x,\bar\zeta_y,\bar\zeta_z)$ for each center in each row.
Since the opposite faces have opposite signs and are degenerate in ranks, we only pick up one of them as in
$\langle\stackrel{\text{4}}{\bullet}\cdot\stackrel{\text{1}}{\bullet}\stackrel{\text{2}}{\bullet}\stackrel{\text{3}}{\bullet}\rangle$,
$\langle\stackrel{\text{2}}{\bullet}\stackrel{\text{3}}{\bullet}\stackrel{\text{4}}{\bullet}\cdot\stackrel{\text{1}}{\bullet}\rangle$,
$\langle\stackrel{\text{1}}{\bullet}\stackrel{\text{3}}{\bullet}\stackrel{\text{4}}{\bullet}\cdot\stackrel{\text{2}}{\bullet}\rangle$,
$\langle\stackrel{\text{1}}{\bullet}\stackrel{\text{2}}{\bullet}\stackrel{\text{4}}{\bullet}\cdot\stackrel{\text{3}}{\bullet}\rangle$,
$\langle\stackrel{\text{1}}{\bullet}\stackrel{\text{4}}{\bullet}\cdot\stackrel{\text{2}}{\bullet}\stackrel{\text{3}}{\bullet}\rangle$,
$\langle\stackrel{\text{2}}{\bullet}\stackrel{\text{4}}{\bullet}\cdot\stackrel{\text{1}}{\bullet}\stackrel{\text{3}}{\bullet}\rangle$,
$\langle\stackrel{\text{3}}{\bullet}\stackrel{\text{4}}{\bullet}\cdot\stackrel{\text{1}}{\bullet}\stackrel{\text{2}}{\bullet}\rangle$.
The rank of this matrix is three and matches the space dimension.
Hence, we conclude that this zonotope is a parallelotope.

\begin{table}[t!]
\begin{center}
\begin{tabular}{r|c|l}
references&$\Delta N$&$\Delta{\bm M}=-2{\bm F}$\\\hline
$\langle\stackrel{\text{1}}{\bullet}\cdot\stackrel{\text{2}}{\bullet}\stackrel{\text{3}}{\bullet}\stackrel{\text{4}}{\bullet}\rangle$&
$-M_1+M_2+M_3+(k_{12}+k_{13}+k_{14})/2$&
$-2({\bm u}_y-{\bm u}_z-{\bm v}_x)$\\
$\langle\stackrel{\text{2}}{\bullet}\cdot\stackrel{\text{1}}{\bullet}\stackrel{\text{3}}{\bullet}\stackrel{\text{4}}{\bullet}\rangle$&
$M_1-M_2+M_3+(k_{12}+k_{23}+k_{24})/2$&
$-2(-{\bm u}_x+{\bm u}_z-{\bm v}_y)$\\
$\langle\stackrel{\text{3}}{\bullet}\cdot\stackrel{\text{1}}{\bullet}\stackrel{\text{2}}{\bullet}\stackrel{\text{4}}{\bullet}\rangle$&
$M_1+M_2-M_3+(k_{13}+k_{23}+k_{34})/2$&
$-2({\bm u}_x-{\bm u}_y-{\bm v}_z)$\\
$\langle\stackrel{\text{4}}{\bullet}\cdot\stackrel{\text{1}}{\bullet}\stackrel{\text{2}}{\bullet}\stackrel{\text{3}}{\bullet}\rangle$&
$-M_1-M_2-M_3+(k_{14}+k_{24}+k_{34})/2$&
$-2({\bm v}_x+{\bm v}_y+{\bm v}_z)$\\
$\langle\stackrel{\text{1}}{\bullet}\stackrel{\text{2}}{\bullet}\cdot\stackrel{\text{3}}{\bullet}\stackrel{\text{4}}{\bullet}\rangle$&
$2M_3+(k_{13}+k_{23}+k_{14}+k_{24})/2$&
$-2(-{\bm u}_x+{\bm u}_y-{\bm v}_x-{\bm v}_y)$\\
$\langle\stackrel{\text{1}}{\bullet}\stackrel{\text{3}}{\bullet}\cdot\stackrel{\text{2}}{\bullet}\stackrel{\text{4}}{\bullet}\rangle$&
$2M_2+(k_{12}+k_{23}+k_{14}+k_{34})/2$&
$-2({\bm u}_x-{\bm u}_z-{\bm v}_x-{\bm v}_z)$\\
$\langle\stackrel{\text{1}}{\bullet}\stackrel{\text{4}}{\bullet}\cdot\stackrel{\text{2}}{\bullet}\stackrel{\text{3}}{\bullet}\rangle$&
$-2M_1+(k_{12}+k_{13}+k_{24}+k_{34})/2$&
$-2({\bm u}_y-{\bm u}_z+{\bm v}_y+{\bm v}_z)$\\
$\langle\stackrel{\text{2}}{\bullet}\stackrel{\text{3}}{\bullet}\cdot\stackrel{\text{1}}{\bullet}\stackrel{\text{4}}{\bullet}\rangle$&
$2M_1+(k_{12}+k_{13}+k_{24}+k_{34})/2$&
$-2(-{\bm u}_y+{\bm u}_z-{\bm v}_y-{\bm v}_z)$\\
$\langle\stackrel{\text{2}}{\bullet}\stackrel{\text{4}}{\bullet}\cdot\stackrel{\text{1}}{\bullet}\stackrel{\text{3}}{\bullet}\rangle$&
$-2M_2+(k_{12}+k_{23}+k_{14}+k_{34})/2$&
$-2(-{\bm u}_x+{\bm u}_z+{\bm v}_x+{\bm v}_z)$\\
$\langle\stackrel{\text{3}}{\bullet}\stackrel{\text{4}}{\bullet}\cdot\stackrel{\text{1}}{\bullet}\stackrel{\text{2}}{\bullet}\rangle$&
$-2M_3+(k_{13}+k_{23}+k_{14}+k_{24})/2$&
$-2({\bm u}_x-{\bm u}_y+{\bm v}_x+{\bm v}_y)$\\
$\langle\stackrel{\text{1}}{\bullet}\stackrel{\text{2}}{\bullet}\stackrel{\text{3}}{\bullet}\cdot\stackrel{\text{4}}{\bullet}\rangle$&
$M_1+M_2+M_3+(k_{14}+k_{24}+k_{34})/2$&
$-2(-{\bm v}_x-{\bm v}_y-{\bm v}_z)$\\
$\langle\stackrel{\text{1}}{\bullet}\stackrel{\text{2}}{\bullet}\stackrel{\text{4}}{\bullet}\cdot\stackrel{\text{3}}{\bullet}\rangle$&
$-M_1-M_2+M_3+(k_{13}+k_{23}+k_{34})/2$&
$-2(-{\bm u}_x+{\bm u}_y+{\bm v}_z)$\\
$\langle\stackrel{\text{1}}{\bullet}\stackrel{\text{3}}{\bullet}\stackrel{\text{4}}{\bullet}\cdot\stackrel{\text{2}}{\bullet}\rangle$&
$-M_1+M_2-M_3+(k_{12}+k_{23}+k_{24})/2$&
$-2({\bm u}_x-{\bm u}_z+{\bm v}_y)$\\
$\langle\stackrel{\text{2}}{\bullet}\stackrel{\text{3}}{\bullet}\stackrel{\text{4}}{\bullet}\cdot\stackrel{\text{1}}{\bullet}\rangle$&
$M_1-M_2-M_3+(k_{12}+k_{13}+k_{14})/2$&
$-2(-{\bm u}_y+{\bm u}_z+{\bm v}_x)$\\
\end{tabular}
\end{center}
\caption{
Changes of the reference rank $\Delta N$ and the relative ranks $\Delta{\bm M}$ when references change in duality cascades.
The translational shifts of the relative ranks $\Delta{\bm M}$ are twice of the center vectors of the faces ${\bm F}$ with negative signs.}
\label{cuboctachange}
\end{table}

Here, besides the pictorial illustration in figure \ref{tilingfigure}, we have confirmed algebraically that the fundamental domain is a parallelotope using the sign matrix $E$ \eqref{fourE}.
Note that the signs for centers are not just an auxiliary mathematical object to examine the zonotope.
In fact, physically, centers of faces encode the information of reference changes in duality cascades.
Namely, as in \cite{FMMN}, when lower ranks appear and references change in duality cascades, the parameters of relative ranks ${\bm M}$ are shifted by translations so that the failure of the inequalities is improved.
Here we study the shifts $\Delta{\bm M}$ in the basis $({\bm u}_x,{\bm u}_y,{\bm u}_z,{\bm v}_x,{\bm v}_y,{\bm v}_z)$ and list them for various reference changes in table \ref{cuboctachange}.
Interestingly, the shifts are nothing but twice of the center vectors with negative signs,
\begin{align}
\Delta{\bm M}=-2{\bm F},
\end{align}
which bring the center (and vertices, respectively) of one face into the center (and vertices) of the opposite face parallelly.
This gives a physical interpretation for the mathematical results on the sign matrix $E$.
Namely, the condition that the rank matches the dimension requires that all of the translations are compatible in the space of the dimension and hence can tile the parameter space by translations.

\subsection{Degeneracies}\label{degenerate}

\begin{figure}[!t]
\centering\includegraphics[scale=0.6,angle=-90]{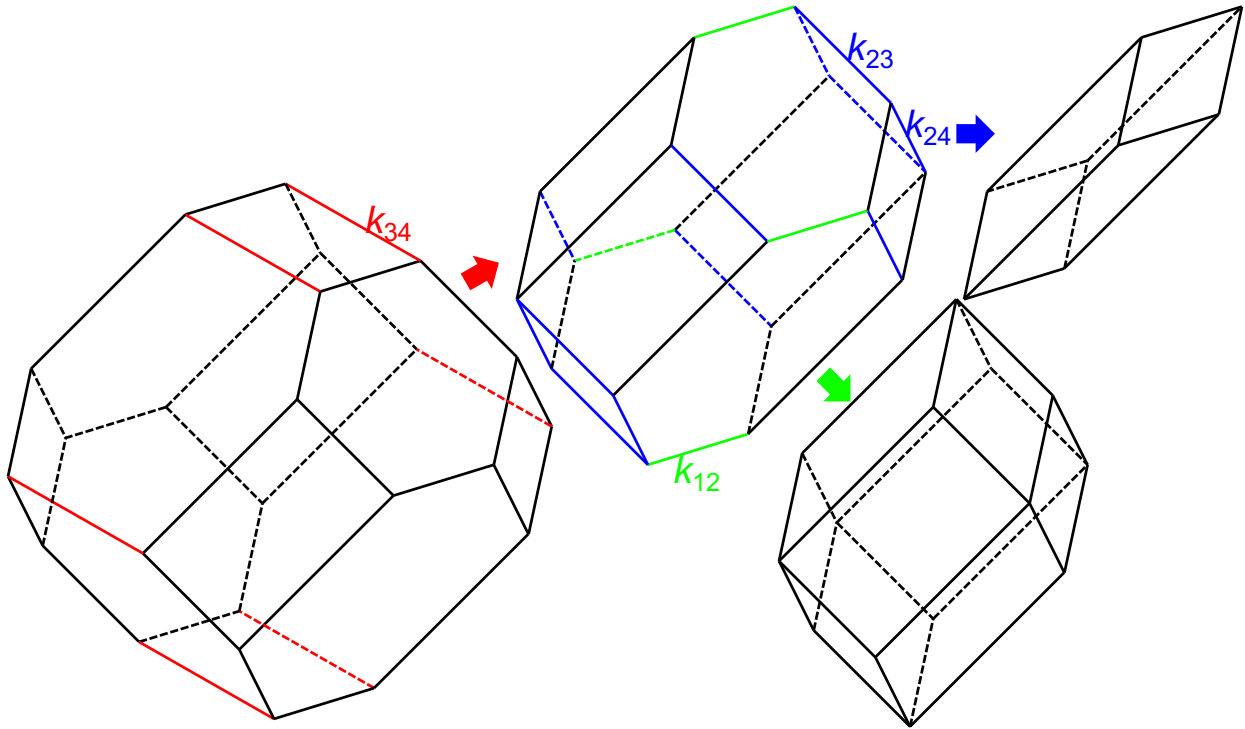}
\caption{Degeneracies from the truncated octahedron.
By contracting one generator corresponding to $k_{34}$, we obtain the elongated dodecahedron.
If we further contract that of $k_{12}$, we obtain the rhombic dodecahedron.
Alternatively, we can contract those of $k_{23}$ and $k_{24}$ and find the cube.}
\label{cuboctadegenerate}
\end{figure}

So far we have considered the most general case with all of four 5-branes different.
Let us turn to degenerate cases where some of 5-branes have identical charges (or multiple of them if we do not care about the stability of 5-branes).
See figure \ref{cuboctadegenerate}.
When $\stackrel{\text{3}}{\bullet}$ and $\stackrel{\text{4}}{\bullet}$ are identical, one of levels vanishes, $k_{34}=0$, and the truncated octahedron degenerates into the elongated dodecahedron (up to combinatorial equivalence; the same applies hereafter).
Furthermore, when $\stackrel{\text{1}}{\bullet}$ and $\stackrel{\text{2}}{\bullet}$ are also identical, another level $k_{12}$ vanishes and the elongated dodecahedron further degenerates into the rhombic dodecahedron studied in \cite{FMMN}.
Alternatively, if $\stackrel{\text{1}}{\bullet}$ and $\stackrel{\text{2}}{\bullet}$ stay different though $\stackrel{\text{2}}{\bullet}$ is identical to $\stackrel{\text{3}}{\bullet}$ and $\stackrel{\text{4}}{\bullet}$, $k_{23}=k_{24}=k_{34}=0$, the elongated dodecahedron degenerates into the cube.

Note that, in three dimensions, the parallelotopes are classified and there are only five types.
In addition to the above four, there is also a hexagonal prism.
This is obtained by setting $k_{24}=k_{34}=0$ while keeping $k_{23}$ non-vanishing.
Although this is possible purely from mathematical viewpoints with $k_{ij}$ independent, for our current system with the levels obtained from the determinants of two 5-brane charges \eqref{leveldet}, the hexagonal prism does not appear.
It is, however, interesting to find how our current system can be generalized to incorporate the hexagonal prism.

To realize the hexagonal prism, we need to keep the HW transitions \eqref{HWpq}, while forget the determinants for $k_{ij}$ \eqref{leveldet}.
This indicates that, although the determinants for $k_{ij}$ \eqref{leveldet} depend on details of brane dynamics, the HW transitions \eqref{HWpq} themselves are obtained from more kinematical relations such as reflections (idempotencies) and charge conservations (Yang-Baxter relations).
To see this, let us assume a linearly generalized expression $s_{ij}$ for the HW transitions exchanging 5-branes $i$ and $j$
\begin{align}
s_{ij}:\cdots K\stackrel{i}{\bullet}L\stackrel{j}{\bullet}M\cdots\to\cdots K\stackrel{j}{\bullet}aK+bL+cM+k_{ij}\stackrel{i}{\bullet}M\cdots,
\end{align}
with coefficients $a,b,c$.
By requiring the idempotency $s_{ij}^2=1$, we find $b=-1$ and $k_{ij}=k_{ji}$, except a few trivial solutions.
Furthermore, we obtain $a=c=1$ by applying the Yang-Baxter relation $s_{12}s_{13}s_{23}=s_{23}s_{13}s_{12}$ to brane configurations $\cdots K\stackrel{1}{\bullet}L\stackrel{2}{\bullet}M\stackrel{3}{\bullet}N\cdots$.
These computations may provide a clue to investigate further the relation between duality cascades and parallelotopes.

In summary, to study our proposal in section \ref{conj} that the fundamental domain defined by inequalities is a parallelotope, we start with brane configurations with four 5-branes of different types in this section.
In addition to the visualization in figure \ref{tilingfigure}, by clarifying the structures, we find that the definition by inequalities can be replaced by the definition by brane configurations without rank differences.
These configurations serve as vertices of the fundamental domain and using these vertices the same fundamental domain can be expressed as a zonotope, where the generators correspond to exchanges of 5-branes.
From this viewpoint we can apply the test of parallelotopes using the sign matrix $E$ \eqref{fourE}.
We have also pointed out that the sign matrix is not just a mathematical object for the test.
Instead, as in table \ref{cuboctachange}, the matrix encodes the information of how duality cascades occur and the condition is equivalent to the compatibility of two different processes of duality cascades.

\section{More 5-branes}\label{more}

In section \ref{conj} we have proposed our conjecture that the fundamental domain is a parallelotope and in the previous section we have studied the conjecture for brane configurations with four 5-branes carefully.
It seems that most of our arguments continue to hold for $d+1$ of 5-branes, which we turn to in this section.
Here we present our arguments for this conjecture.
To avoid our arguments from being too abstract, we provide various examples from the previous section when necessary.

\subsection{Equivalence}\label{equiv}

First let us introduce several descriptions for the fundamental domain by generalizing the previous studies and argue their equivalence abstractly.
As previously we first assume that 5-branes are all different and only comment on degenerate cases with some of them identical by contractions at the end.

If we fix an order of 5-branes as the standard order, we can always bring 5-branes in brane configurations into the standard order by the HW transitions and read off the relative ranks.
In this sense we switch freely between brane configurations and relative ranks in a fixed order in the following.

{\it The ${\cal H}$ description.}
The first description of the fundamental domain is by duality cascades.
We have defined the fundamental domain to be obtained as the final destinations of duality cascades, where duality cascades occur no more; in other words, no lower ranks compared with the reference rank appear in applying the HW transitions arbitrarily without crossing the reference.
If we parameterize the relative ranks in $d$ intervals by ${\bm M}=(M_i)_{i=1}^d$, the fundamental domain is given schematically by a system of linear inequalities
\begin{align}
{\cal H}=\{{\bm M}\in{\mathbb R}^d\;|\;{\bm a}_f\cdot{\bm M}\le b_f\;(1\le f\le 2^{d+1}-2)\},
\label{facets}
\end{align}
with coefficients ${\bm a}_f$, $b_f$.
This corresponds to \eqref{cuboctaineq} for the previous case with four 5-branes.
Note that, as below \eqref{cuboctaineq}, the inequalities are labeled by how we split 5-branes into two disjoint groups.
For the case with four 5-branes, we have seen the correspondence between the splits and the inequalities in table \ref{cuboctainequalities}.
Since each inequality restricts to a half space and defines a facet (a face of codimension 1)\footnote{For a polytope in higher dimensions, usually the word ``faces'' is reserved for its simplices of all dimensions, including vertices, edges, facets and so on.}, the entire system of inequalities defines a polytope.
Besides, from the definition with inequalities, apparently ${\cal H}$ is convex.
From the construction, the elements of ${\cal H}$ can be judged by whether no lower ranks than the reference appear in applying the HW transitions arbitrarily without crossing the reference.

\begin{figure}[t!]
\centering\includegraphics[scale=0.6,angle=-90]{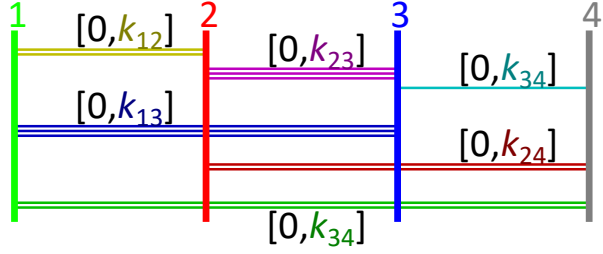}
\caption{Brane configurations satisfying the S-rule.
The number of D3-branes stretching between two 5-branes does not exceed the level $k_{ij}$ determined by the two 5-branes in \eqref{leveldet}.}
\label{Srule}
\end{figure}

\begin{figure}[t!]
\centering\includegraphics[scale=0.6,angle=-90]{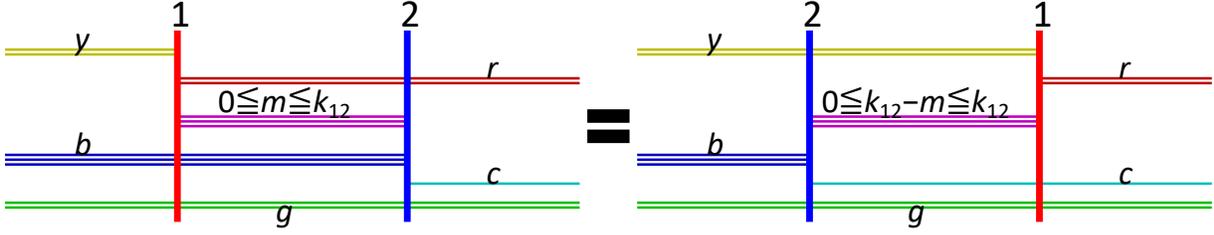}
\caption{Brane configurations obtained in applying the HW transitions.
If a configuration satisfies the S-rule, so does the configuration obtained in the HW transitions.
Indeed, let us consider the brane configuration corresponding to figure \ref{cartoon} and indicate the number of D3-branes in each stack by the initial letter of its color (such as red/green/blue and cyan/magenta/yellow).
Then, in the brane configuration before the HW transition (left) we have $K=y+b+g$, $L=r+m+b+g$, $M=r+c+g$, while in that after the HW transition (right) the induced total rank $K-L+M+k_{12}$ in \eqref{HWpq} can be assigned to each stack with the numbers in $[0,k_{ij}]$ by $K-L+M+k_{12}=y+(k_{12}-m)+c+g$.}
\label{SruleHW}
\end{figure}

{\it The ${\cal Z}$ description.}
The second description is by the so-called S-rule \cite{HW}.
It was argued that supersymmetric brane configurations can be constructed by restricting the number of D3-branes $M_{ij}$ stretching between each pairs of 5-branes labelled by $(i,j)$ not to exceed $k_{ij}$, $M_{ij}\in[0,k_{ij}]$ (see figure \ref{Srule}).
Note that this condition is invariant under the HW transitions and does not depend on the order of 5-branes (see figure \ref{SruleHW}).
By identifying these brane configurations in the parameter space of relative ranks, the number $M_{ij}$ actually defines a vector ${\bm M}_{ij}$ in ${\mathbb R}^d$, where the direction is specified by changing each $M_{ij}$ in $[0,k_{ij}]$ with the others kept fixed.
Hence, by collecting all of the vectors, we consider the subspace in ${\mathbb R}^d$ denoted schematically as
\begin{align}
{\cal Z}=\Big\{\sum_{(i,j)}^{d(d+1)/2}{\bm M}_{ij}\;\Big|\;||{\bm M_{ij}}||\le k_{ij}\Big\}.
\label{Zdescription}
\end{align}
where the directions of ${\bm M}_{ij}$ are fixed with the norm defined suitably and the summation runs over all the $d(d+1)/2$ pairs of 5-branes.
For the previous case with four 5-branes, this description corresponds to \eqref{Msigns} with the norms of coefficients $(\epsilon_x,\epsilon_y,\epsilon_z,\zeta_x,\zeta_y,\zeta_z)$ not larger than $1$.
There, we have also found that the vectors correspond to the exchanges of two 5-branes in \eqref{commutation}.
Mathematically, this is the Minkowski (vector) sum of vectors known as a zonotope (see e.g. \cite{polytope}) and can be regarded as a projection from a $d(d+1)/2$-dimensional hyperrectangle, which is apparently convex.
It might be confusing from physical viewpoints that the S-rule should allow all physically possible brane configurations instead of only the final destinations of duality cascades.
As pointed out in \cite{EK}, for those not located in the fundamental domain, the brane configurations satisfy the S-rule by allowing D3-branes to connect 5-branes across the reference.

{\it The ${\cal V}$ description.}
The third description is by brane configurations without rank differences but in all orders of 5-branes, which are schematically denoted as ${\langle}{\bullet}{\bullet}{\cdots}{\bullet}{\rangle}$.
As noted previously, after fixing the standard order, we can bring 5-branes into the standard order by applying the HW transitions, read off the relative ranks ${\bm M}$ and plot them in the parameter space ${\mathbb R}^d$.
Let us define ${\cal V}$ to be the convex hull of these points and indicate it schematically by
\begin{align}
{\cal V}=\Big\{\sum_{v=1}^{(d+1)!}\lambda_v{\langle}{\bullet}{\bullet}{\cdots}{\bullet}{\rangle}_v\;\Big|\;\sum_v\lambda_v=1,\;\lambda_v\ge 0\Big\},
\end{align}
where $v$ runs over $(d+1)!$ orders of 5-branes, such as $\langle\stackrel{\text{3}}{\bullet}\stackrel{\text{1}}{\bullet}\stackrel{\text{4}}{\bullet}\stackrel{\text{2}}{\bullet}\rangle$ in table \ref{cuboctasigns} for the case with four 5-branes.

{\it Equivalence.}
In the following we show abstractly the equivalence of these three descriptions,
\begin{align}
{\cal V}={\cal Z}={\cal H},
\end{align}
for brane configurations with arbitrary numbers of 5-branes.

First note that, since the brane configurations without rank differences ${\langle}{\bullet}{\bullet}{\cdots}{\bullet}{\rangle}$ appearing in ${\cal V}$ are trivially elements of ${\cal Z}$ by setting all of the parameters $M_{ij}$ to be zero and ${\cal V}$ is defined by their convex hull, ${\cal V}$ is a subset of ${\cal Z}$, ${\cal V}\subset{\cal Z}$.

Next, since the condition for ${\cal Z}$ is invariant under the HW transitions (see figure \ref{SruleHW}), the configurations of ${\cal Z}$ in any order of 5-branes with $M_{ij}\in[0,k_{ij}]$ apparently give configurations with no lower ranks in all intervals.
For this reason, ${\cal Z}$ is a subset of ${\cal H}$, ${\cal Z}\subset{\cal H}$.

Now, let us claim that the brane configurations without rank differences ${\langle}{\bullet}{\bullet}{\cdots}{\bullet}{\rangle}$ appearing in the definition of ${\cal V}$ are always vertices of ${\cal H}$.
First note that vertices of a polytope defined from inequalities are the points satisfying as many inequalities marginally as the dimension $d$ to evade from the polytope.
For the brane configurations without rank differences ${\langle}{\bullet}{\bullet}{\cdots}{\bullet}{\rangle}$, there are $d$ inequalities coming from each interval of 5-branes in the fixed order.
Hence, we conclude that the brane configurations without rank differences ${\langle}{\bullet}{\bullet}{\cdots}{\bullet}{\rangle}$ are vertices of ${\cal H}$.

For example, for the case with four 5-branes, the brane configuration $\langle\stackrel{\text{3}}{\bullet}\stackrel{\text{1}}{\bullet}\stackrel{\text{4}}{\bullet}\stackrel{\text{2}}{\bullet}\rangle$ (or the relative ranks in the standard order of 5-branes) satisfies marginally the three inequalities labelled by $\langle\stackrel{\text{3}}{\bullet}\cdot\stackrel{\text{1}}{\bullet}\stackrel{\text{4}}{\bullet}\stackrel{\text{2}}{\bullet}\rangle$, $\langle\stackrel{\text{3}}{\bullet}\stackrel{\text{1}}{\bullet}\cdot\stackrel{\text{4}}{\bullet}\stackrel{\text{2}}{\bullet}\rangle$, $\langle\stackrel{\text{3}}{\bullet}\stackrel{\text{1}}{\bullet}\stackrel{\text{4}}{\bullet}\cdot\stackrel{\text{2}}{\bullet}\rangle$ and serves as a vertex for ${\cal H}$.

Next we argue that ${\cal H}$ has no other vertices.
For this purpose, let us refer to two facets (labeled by splitting 5-branes, as noted below \eqref{facets} or below \eqref{cuboctaineq}) as {\it compatible}, if there is an inclusion relation for the combinations of 5-branes between the reference rank and the rank denoted by $\cdot$.
Note that later we claim that {\it two hyperplanes of facets intersect on the polytope only when they are compatible}.
Then, for vertices (with $d$ hyperplanes of facets intersecting on the polytope), all pairs of $d$ facets have to be compatible and the partial order set defined with the inclusion relations turns out to be a total order set.
For this reason, all ranks in the intervals are vanishing and the vertices of ${\cal H}$ have to be the brane configurations without rank differences ${\langle}{\bullet}{\bullet}{\cdots}{\bullet}{\rangle}$.

Again, let us revisit the argument for the case with four 5-branes.
$\langle\stackrel{\text{1}}{\bullet}\cdot\stackrel{\text{2}}{\bullet}\stackrel{\text{3}}{\bullet}\stackrel{\text{4}}{\bullet}\rangle$ is compatible with $\langle\stackrel{\text{1}}{\bullet}\stackrel{\text{3}}{\bullet}\cdot\stackrel{\text{2}}{\bullet}\stackrel{\text{4}}{\bullet}\rangle$ since $\{1\}\subset\{1,3\}$, while not compatible with $\langle\stackrel{\text{2}}{\bullet}\cdot\stackrel{\text{1}}{\bullet}\stackrel{\text{3}}{\bullet}\stackrel{\text{4}}{\bullet}\rangle$ since there are no inclusion relations between $\{1\}$ and $\{2\}$.
As pointed out, the intersection on the polytope between two hyperplanes is characterized from the compatibility.
Namely, the facet $\langle\stackrel{\text{1}}{\bullet}\cdot\stackrel{\text{2}}{\bullet}\stackrel{\text{3}}{\bullet}\stackrel{\text{4}}{\bullet}\rangle$ shares with the (compatible) facet $\langle\stackrel{\text{1}}{\bullet}\stackrel{\text{3}}{\bullet}\cdot\stackrel{\text{2}}{\bullet}\stackrel{\text{4}}{\bullet}\rangle$ the edge connecting $\langle\stackrel{\text{1}}{\bullet}\stackrel{\text{3}}{\bullet}\stackrel{\text{2}}{\bullet}\stackrel{\text{4}}{\bullet}\rangle$ and $\langle\stackrel{\text{1}}{\bullet}\stackrel{\text{3}}{\bullet}\stackrel{\text{4}}{\bullet}\stackrel{\text{2}}{\bullet}\rangle$, while do not intersect on the polytope with the (non-compatible) facet $\langle\stackrel{\text{2}}{\bullet}\cdot\stackrel{\text{1}}{\bullet}\stackrel{\text{3}}{\bullet}\stackrel{\text{4}}{\bullet}\rangle$ (as seen from figure \ref{cuboctafigure}).
Since vertices are obtained only when three facets intersect on the polytope, these three facets have to be all compatible.
This requires the inclusion relation to form a total order set, such as $(\emptyset\subset)\{3\}\subset\{3,1\}\subset\{3,1,4\}(\subset\{3,1,4,2\})$, which corresponds to the combination of three facets with labels $\langle\stackrel{\text{3}}{\bullet}\cdot\stackrel{\text{1}}{\bullet}\stackrel{\text{4}}{\bullet}\stackrel{\text{2}}{\bullet}\rangle$, $\langle\stackrel{\text{3}}{\bullet}\stackrel{\text{1}}{\bullet}\cdot\stackrel{\text{4}}{\bullet}\stackrel{\text{2}}{\bullet}\rangle$ and $\langle\stackrel{\text{3}}{\bullet}\stackrel{\text{1}}{\bullet}\stackrel{\text{4}}{\bullet}\cdot\stackrel{\text{2}}{\bullet}\rangle$.
In this case, for the vertex (the intersection of the three facets), each interval has to avoid rank differences.
Hence, the vertex has to be the brane configuration without rank differences $\langle\stackrel{\text{3}}{\bullet}\stackrel{\text{1}}{\bullet}\stackrel{\text{4}}{\bullet}\stackrel{\text{2}}{\bullet}\rangle$.

To summarize, we find the inclusion relation ${\cal V}\subset{\cal Z}\subset{\cal H}$ and vertices of ${\cal H}$ are the brane configurations without rank differences ${\langle}{\bullet}{\bullet}{\cdots}{\bullet}{\rangle}$ appearing in ${\cal V}$.
From this, we conclude the equivalence of the three descriptions, ${\cal V}={\cal Z}={\cal H}$.

Let us return to argue the claim that {\it two hyperplanes of facets intersect on the polytope only when they are compatible}.
We argue this claim by explaining that, when configurations in the fundamental domain are on one hyperplane, they are never located on another non-compatible hyperplane by following the HW transitions.
Namely, for example for the case with four 5-branes, let us consider configurations on one facet of the fundamental domain which satisfy one inequality marginally such as $\langle\stackrel{\text{1}}{\bullet}\cdot\stackrel{\text{2}}{\bullet}\stackrel{\text{3}}{\bullet}\stackrel{\text{4}}{\bullet}\rangle$.
Although these configurations are possibly located on another compatible hyperplane such as $\langle\stackrel{\text{1}}{\bullet}\stackrel{\text{3}}{\bullet}\cdot\stackrel{\text{2}}{\bullet}\stackrel{\text{4}}{\bullet}\rangle$, in the following we argue that they can never be on non-compatible hyperplanes such as $\langle\stackrel{\text{2}}{\bullet}\cdot\stackrel{\text{1}}{\bullet}\stackrel{\text{3}}{\bullet}\stackrel{\text{4}}{\bullet}\rangle$.

Indeed, let us consider configurations on one facet of the fundamental domain $\langle\cdots\stackrel{3}{\bullet}\stackrel{2}{\bullet}\stackrel{1}{\bullet}\cdot\stackrel{1'}{\bullet}\stackrel{2'}{\bullet}\stackrel{3'}{\bullet}\cdots\rangle$ or $\langle\cdots\stackrel{3}{\bullet}M_2\stackrel{2}{\bullet}M_1\stackrel{1}{\bullet}0\stackrel{1'}{\bullet}M'_1\stackrel{2'}{\bullet}M'_2\stackrel{3'}{\bullet}\cdots\rangle$ with relative ranks specified.
Though the orders of 5-branes before and after the interval $\cdots\stackrel{1}{\bullet}0\stackrel{1'}{\bullet}\cdots$ are irrelevant, for explanation let us rearrange the orders and relabel 5-branes to be removed as $\stackrel{1}{\bullet},\stackrel{2}{\bullet},\cdots$ and those to be added as $\stackrel{1'}{\bullet},\stackrel{2'}{\bullet},\cdots$ in the non-compatible hyperplane.
Then, as is clear from the sample computation
\begin{align}
&\langle\cdots\stackrel{3}{\bullet}M_2\stackrel{2}{\bullet}M_1\stackrel{1}{\bullet}0\stackrel{1'}{\bullet}M'_1\stackrel{2'}{\bullet}M'_2\stackrel{3'}{\bullet}\cdots\rangle\nonumber\\
&=\langle\cdots\stackrel{3}{\bullet}M_2
\stackrel{1'}{\bullet}M_2+M'_1\;\begin{matrix}+\;k_{11'}\\+\;k_{21'}\end{matrix}
\stackrel{2'}{\bullet}M_2+M'_2\;\begin{matrix}+\;k_{11'}+k_{12'}\\+\;k_{21'}+k_{22'}\end{matrix}
\stackrel{2}{\bullet}M_1+M'_2\;\begin{matrix}+\;k_{11'}+k_{12'}\end{matrix}
\stackrel{1}{\bullet}M_2
\stackrel{3'}{\bullet}\cdots\rangle,
\end{align}
(with the summands put in the matrix form), the interval ranks of 5-branes non-compatible with the original one $\langle\cdots\stackrel{3}{\bullet}\stackrel{2}{\bullet}\stackrel{1}{\bullet}\cdot\stackrel{1'}{\bullet}\stackrel{2'}{\bullet}\stackrel{3'}{\bullet}\cdots\rangle$, such as $\langle\cdots\stackrel{3}{\bullet}\stackrel{1'}{\bullet}\cdot\stackrel{2}{\bullet}\stackrel{1}{\bullet}\stackrel{2'}{\bullet}\stackrel{3'}{\bullet}\cdots\rangle$, $\langle\cdots\stackrel{3}{\bullet}\stackrel{1'}{\bullet}\stackrel{2'}{\bullet}\cdot\stackrel{2}{\bullet}\stackrel{1}{\bullet}\stackrel{3'}{\bullet}\cdots\rangle$ and $\langle\cdots\stackrel{3}{\bullet}\stackrel{2}{\bullet}\stackrel{1'}{\bullet}\stackrel{2'}{\bullet}\cdot\stackrel{1}{\bullet}\stackrel{3'}{\bullet}\cdots\rangle$, have to be strictly larger than the reference rank (because of $k_{ij'}>0$).
For this reason, brane configurations cannot simultaneously satisfy both of the inequalities $\langle\cdots\stackrel{3}{\bullet}\stackrel{2}{\bullet}\stackrel{1}{\bullet}\cdot\stackrel{1'}{\bullet}\stackrel{2'}{\bullet}\stackrel{3'}{\bullet}\cdots\rangle$ and $\langle\cdots\stackrel{3}{\bullet}\stackrel{1'}{\bullet}\cdot\stackrel{2}{\bullet}\stackrel{1}{\bullet}\stackrel{2'}{\bullet}\stackrel{3'}{\bullet}\cdots\rangle$ (or $\langle\cdots\stackrel{3}{\bullet}\stackrel{2}{\bullet}\stackrel{1'}{\bullet}\stackrel{2'}{\bullet}\cdot\stackrel{1}{\bullet}\stackrel{3'}{\bullet}\cdots\rangle$) marginally, for example.

As a final remark, at the end of section \ref{descriptions}, we have found that, if we fix the interval splitting 5-branes (facets), vertices appearing on the facet are direct products of permutations of 5-branes before the interval and those after the interval.
Also, if we fix the order of 5-branes (vertices), the inequalities coming from the intervals in the order form a cone with the brane configuration without rank differences at the tip.

\subsection{Parallelotope}

In the previous subsection, we have presented the three descriptions for the fundamental domain.
Especially, we rewrite the ${\cal H}$ description originating directly from duality cascades into the ${\cal Z}$ description by zonotopes, where we can discuss the fundamental domain with nice properties of zonotopes.

Let us label $d+1$ 5-branes by $C=\{1,2,\cdots,d+1\}$ and fully take the advantage of the ${\cal Z}$ description by zonotopes.
As discussed around \eqref{Zdescription}, for each pair of 5-branes $(i,j)$ with $i<j$, we can stretch at most $k_{ij}(\ge 0)$ of D3-branes between them, which associates to a generator exchanging the two 5-branes.
For this reason, we can define the generator $2{\bm v}_{j,i}$ $(i<j)$ to be $2{\bm v}_{j,i}=[\stackrel{j}{\bullet},\stackrel{i}{\bullet}]$ or
\begin{align}
{\bm v}_{j,i}
=\frac{k_{ij}}{2}(0\stackrel{1}{\bullet}\cdots\stackrel{i-1}{\bullet}0\stackrel{i}{\bullet}
1\stackrel{i+1}{\bullet}1\stackrel{i+2}{\bullet}\cdots\stackrel{j-1}{\bullet}1\stackrel{j}{\bullet}
0\stackrel{j+1}{\bullet}\cdots\stackrel{d+1}{\bullet}),
\end{align}
where we have adopted the notation introduced in \eqref{commutation} by expressing the exchanges of 5-branes with commutation relations and regarding the array of relative ranks directly as a vector.
For the reversed order in indices $i>j$, we reverse the orientation ${\bm v}_{j,i}=-{\bm v}_{i,j}$.
Then, the fundamental domain is a zonotope generated by these vectors $2{\bm v}_{j,i}$ and each vertex can be expressed as
\begin{align}
{\bm V}=\sum_{i<j}\epsilon_{ji}{\bm v}_{j,i},
\end{align}
with the signs $\epsilon_{ji}=\pm 1$.

As in section \ref{signmattest}, with the description by zonotopes, it is easy to test whether the polytope can tile the whole space by studying whether the rank of the sign matrix for centers of facets $E$ \eqref{fourE} coincides with the space dimension, where the center of each facet is obtained by averaging all vertices on it.
Physically, as explained at the end of section \ref{signmattest}, this is nothing but the condition that two processes of duality cascades are compatible with each other.

We would first like to sketch the argument that the condition holds generally for arbitrary numbers of 5-branes.
In averaging all vertices on each facet in determining the sign matrix, those generators ${\bm v}_{j,i}$ connecting 5-branes within the two groups before or after the interval are averaged to zero and only generators connecting between the two groups have non-vanishing sign coefficients.
Since these coefficients have correct signs respecting the orientations, they can be generated from generators starting from each 5-brane, whose number is identical to the space dimension $d$.
Hence, the fundamental domain for 5-branes of different types is a parallelotope.

More explicitly, the center of a facet is given generally by
\begin{align}
{\bm F}=\sum_{i<j}\bar\epsilon_{ji}{\bm v}_{j,i},
\label{cvector}
\end{align}
for zonotopes, where $\bar\epsilon_{ji}$ takes either signs or zeros.
From our setup of brane configurations, we can fix $\bar\epsilon_{ji}$ as follows.
First note that facets are labeled by splitting 5-branes $C$ into two disjoint sets $C=A\sqcup B$.
Then, vertices on the facet $\langle A\cdot B\rangle$ are given by direct products of all permutations of $A=\{a_1,a_2,\cdots,a_\alpha\}$ and all permutations of $B=\{b_1,b_2,\cdots,b_\beta\}$ with $\alpha+\beta=d+1$,
\begin{align}
\langle\underbrace{\stackrel{a_1}{\bullet}\stackrel{a_2}{\bullet}\cdots\stackrel{a_\alpha}{\bullet}}_{\text{permutations}}\underbrace{\stackrel{b_1}{\bullet}\stackrel{b_2}{\bullet}\cdots\stackrel{b_\beta}{\bullet}}_{\text{ permutations}}\rangle.
\label{permvertex}
\end{align}
By taking the average for the vertices on the facet,
\begin{align}
{\bm F}_{\langle A\cdot B\rangle}=\frac{1}{\alpha!\beta!}\sum_{\langle\bullet\bullet\cdots\bullet\rangle\in\langle A\cdot B\rangle}
{\bm V}_{\langle\underbrace{\bullet\bullet\cdots\bullet}_{A}\underbrace{\bullet\bullet\cdots\bullet}_{B}\rangle},
\end{align}
where the sum is taken over all the $\alpha!\beta!$ vertices \eqref{permvertex} on the facet $\langle A\cdot B\rangle$, it is clear that generators ${\bm v}_{j,i}$ connecting those within $A$ or those within $B$ are cancelled and only those connecting between $A$ and $B$ remain with correct signs
\begin{align}
{\bm F}_{\langle A\cdot B\rangle}=-\sum_{(a,b)\in A\times B}{\bm v}_{b,a}.
\label{cv}
\end{align}

If we choose
\begin{align}
{\bm F}_i={\bm F}_{\langle\{i\}\cdot C\backslash\{i\}\rangle}=-\sum_{j\ne i}{\bm v}_{j,i},
\label{cbasis}
\end{align}
$(1\le i\le d+1)$ as a basis subject to one constraint $\sum_{i=1}^{d+1}{\bm F}_i={\bm 0}$, it is not difficult to find by induction that all the remaining centers are their linear combinations,
\begin{align}
{\bm F}_{\langle A\cdot B\rangle}=\sum_{a\in A}{\bm F}_a.
\end{align}
Since we can express the centers with $\{{\bm F}_i\}_{i=1}^d$ by eliminating ${\bm F}_{d+1}=-\sum_{i=1}^d{\bm F}_i$, the rank of the matrix $E$ defined by signs coincides with the space dimension $d$, which implies that the zonotope tiles the space.

In section \ref{signmattest} for the case with four 5-branes, we have studied the four vertices on the facet $\langle\stackrel{\text{1}}{\bullet}\stackrel{\text{3}}{\bullet}\cdot\stackrel{\text{2}}{\bullet}\stackrel{\text{4}}{\bullet}\rangle$ and their average below \eqref{centersign} (see also table \ref{cuboctasigns} and table \ref{cuboctafaces}).
Let us reconsider the same example in our current notation with $A=\{1,3\}$ and $B=\{2,4\}$.
Note that, although the previous notation is suitable for visualization in three dimensions, the current one works better for general dimensions.
The four vertices are $-{\bm v}_{2,1}\pm{\bm v}_{3,1}-{\bm v}_{4,1}+{\bm v}_{3,2}\pm{\bm v}_{4,2}-{\bm v}_{4,3}$ and their average is ${\bm F}=-{\bm v}_{2,1}-{\bm v}_{4,1}+{\bm v}_{3,2}-{\bm v}_{4,3}$.
Here note that signs between $\stackrel{\text{1}}{\bullet}$ and $\stackrel{\text{3}}{\bullet}$ in $A$ (or $\stackrel{\text{2}}{\bullet}$ and $\stackrel{\text{4}}{\bullet}$ in $B$ respectively) cancel among themselves and leave only generators connecting $\stackrel{\text{1}}{\bullet}$/$\stackrel{\text{3}}{\bullet}$ and $\stackrel{\text{2}}{\bullet}$/$\stackrel{\text{4}}{\bullet}$.
By preparing ${\bm F}_1=-{\bm v}_{2,1}-{\bm v}_{3,1}-{\bm v}_{4,1}$, ${\bm F}_2={\bm v}_{2,1}-{\bm v}_{3,2}-{\bm v}_{4,2}$, ${\bm F}_3={\bm v}_{3,1}+{\bm v}_{3,2}-{\bm v}_{4,3}$, ${\bm F}_4={\bm v}_{4,1}+{\bm v}_{4,2}+{\bm v}_{4,3}$ subject to one constraint ${\bm F}_1+{\bm F}_2+{\bm F}_3+{\bm F}_4=0$, we can express the result with them as ${\bm F}={\bm F}_1+{\bm F}_3$.
Note that the unpleasant terms $\pm{\bm v}_{3,1}$ within $A$ cancel among themselves.
In table \ref{facetsign}, we repeat the same computation as table \ref{cuboctafaces}.

\begin{table}[t!]
\begin{center}
\begin{tabular}{c|c|l}
facets&$(\bar\epsilon_{21},\bar\epsilon_{31},\bar\epsilon_{41},\bar\epsilon_{32},\bar\epsilon_{42},\bar\epsilon_{43})$&${\bm F}$\\\hline
$\langle\stackrel{\text{1}}{\bullet}\cdot\stackrel{\text{2}}{\bullet}\stackrel{\text{3}}{\bullet}\stackrel{\text{4}}{\bullet}\rangle$&
$(-,-,-,\;0,\;0,\;0)$&${\bm F}_1$\\
$\langle\stackrel{\text{2}}{\bullet}\cdot\stackrel{\text{1}}{\bullet}\stackrel{\text{3}}{\bullet}\stackrel{\text{4}}{\bullet}\rangle$&
$(+,\;0,\;0,-,-,\;0)$&${\bm F}_2$\\
$\langle\stackrel{\text{3}}{\bullet}\cdot\stackrel{\text{1}}{\bullet}\stackrel{\text{2}}{\bullet}\stackrel{\text{4}}{\bullet}\rangle$&
$(\;0,+,\;0,+,\;0,-)$&${\bm F}_3$\\
$\langle\stackrel{\text{4}}{\bullet}\cdot\stackrel{\text{1}}{\bullet}\stackrel{\text{2}}{\bullet}\stackrel{\text{3}}{\bullet}\rangle$&
$(\;0,\;0,+,\;0,+,+)$&${\bm F}_4=-{\bm F}_1-{\bm F}_2-{\bm F}_3$\\
$\langle\stackrel{\text{1}}{\bullet}\stackrel{\text{2}}{\bullet}\cdot\stackrel{\text{3}}{\bullet}\stackrel{\text{4}}{\bullet}\rangle$&
$(\;0,-,-,-,-,\;0)$&${\bm F}_1+{\bm F}_2$\\
$\langle\stackrel{\text{1}}{\bullet}\stackrel{\text{3}}{\bullet}\cdot\stackrel{\text{2}}{\bullet}\stackrel{\text{4}}{\bullet}\rangle$&
$(-,\;0,-,+,\;0,-)$&${\bm F}_1+{\bm F}_3$\\
$\langle\stackrel{\text{1}}{\bullet}\stackrel{\text{4}}{\bullet}\cdot\stackrel{\text{2}}{\bullet}\stackrel{\text{3}}{\bullet}\rangle$&
$(-,-,\;0,\;0,+,+)$&${\bm F}_1+{\bm F}_4=-{\bm F}_2-{\bm F}_3$\\
$\langle\stackrel{\text{2}}{\bullet}\stackrel{\text{3}}{\bullet}\cdot\stackrel{\text{1}}{\bullet}\stackrel{\text{4}}{\bullet}\rangle$&
$(+,+,\;0,\;0,-,-)$&${\bm F}_2+{\bm F}_3$\\
$\langle\stackrel{\text{2}}{\bullet}\stackrel{\text{4}}{\bullet}\cdot\stackrel{\text{1}}{\bullet}\stackrel{\text{3}}{\bullet}\rangle$&
$(+,\;0,+,-,\;0,+)$&${\bm F}_2+{\bm F}_4=-{\bm F}_1-{\bm F}_3$\\
$\langle\stackrel{\text{3}}{\bullet}\stackrel{\text{4}}{\bullet}\cdot\stackrel{\text{1}}{\bullet}\stackrel{\text{2}}{\bullet}\rangle$&
$(\;0,+,+,+,+,\;0)$&${\bm F}_3+{\bm F}_4=-{\bm F}_1-{\bm F}_2$\\
$\langle\stackrel{\text{1}}{\bullet}\stackrel{\text{2}}{\bullet}\stackrel{\text{3}}{\bullet}\cdot\stackrel{\text{4}}{\bullet}\rangle$&
$(\;0,\;0,-,\;0,-,-)$&${\bm F}_1+{\bm F}_2+{\bm F}_3$\\
$\langle\stackrel{\text{1}}{\bullet}\stackrel{\text{2}}{\bullet}\stackrel{\text{4}}{\bullet}\cdot\stackrel{\text{3}}{\bullet}\rangle$&
$(\;0,-,\;0,-,\;0,+)$&${\bm F}_1+{\bm F}_2+{\bm F}_4=-{\bm F}_3$\\
$\langle\stackrel{\text{1}}{\bullet}\stackrel{\text{3}}{\bullet}\stackrel{\text{4}}{\bullet}\cdot\stackrel{\text{2}}{\bullet}\rangle$&
$(-,\;0,\;0,+,+,\;0)$&${\bm F}_1+{\bm F}_3+{\bm F}_4=-{\bm F}_2$\\
$\langle\stackrel{\text{2}}{\bullet}\stackrel{\text{3}}{\bullet}\stackrel{\text{4}}{\bullet}\cdot\stackrel{\text{1}}{\bullet}\rangle$&
$(+,+,+,\;0,\;0,\;0)$&${\bm F}_2+{\bm F}_3+{\bm F}_4=-{\bm F}_1$
\end{tabular}
\end{center}
\caption{Centers of facets.
We repeat the analysis in table \ref{cuboctafaces} in a different notation with \eqref{cvector} and \eqref{cbasis}, where the four center vectors ${\bm F}_{i=1,\cdots,4}$ are subject to the constraint $\sum_{i=1}^4{\bm F}_i=0$.}
\label{facetsign}
\end{table}

As pointed out in section \ref{descriptions}, the fundamental domain in figure \ref{cuboctafigure} is combinatorially equivalent to figure 15.1.5 in \cite{handbook}.
In addition to zonotopes, the figure in \cite{handbook} also explains the so-called permutohedron in three dimensions, which is the polytope with vertices located at the coordinate $(1,2,3,4)$ and its permutations.
It is natural to expect that the fundamental domain with more 5-branes discussed in this section continue to be combinatorially equivalent to permutohedrons of higher dimensions.

For brane configurations with some of 5-branes identical, all we have to do is to take the degenerate limits by contracting the zonotope.
As is clear by taking the norms of the generators to be infinitesimal, if a zonotope tiles the space, so do its contractions as long as the dimensionality is kept fixed.
For this reason, we claim that brane configurations with 5-branes of arbitrary types always give rise to parallelotopes.

To summarize, by generalizing previous studies with four 5-branes to arbitrary numbers, we have found that the fundamental domain is always a parallelotope.
The main argument is to rewrite the fundamental domain in the ${\cal H}$ description into the ${\cal Z}$ description.
Then, by utilizing the test for zonotopes, the proposal that the polytope is a parallelotope is guaranteed from combinatorial properties of zonotopes.

\section{Conclusions}

Starting from the physical questions on the working hypothesis of duality cascades in section \ref{dc}, whether duality cascades always end and whether the final destination of duality cascades is unique regardless of processes, we have found deep connections to parallelotopes.
Namely, since duality cascades are realized by translations, the questions are reformulated into whether the fundamental domain (of supersymmetric brane configurations in duality cascades) is a parallelotope.
In this paper, we apply theories of combinatorial geometries and answer this question positively.
Let us list several directions we wish to pursue in the future.

Firstly, note that although our previous studies in \cite{FMMN} confine to the cases of quantum curves with symmetries of Weyl groups, the question was answered by discovering hidden symmetries of affine Lie algebras.
In contrast, our current studies are applicable generally to brane configurations with arbitrary combinations of 5-branes.
The viewpoint of symmetries, however, is missing.
It is interesting to revisit the fundamental domain studied in this paper from the viewpoint of symmetries.

Secondly,  parallelotopes in three dimensions are classified into five.
As pointed out in section \ref{degenerate}, only four of them appear in our current setup of brane configurations and the hexagonal prism is missing.
We would like to find out its realization by deforming brane configurations.

Thirdly, the description of the fundamental domain by zonotopes indicates the projection from a higher-dimensional hyperrectangle.
Physically, this implies that we do not distinguish D3-branes disconnected by the middle 5-branes and those bypassing the middle 5-branes in brane configurations as long as the total sum in each interval is identical.
It is interesting to distinguish them with physical observables.

Fourthly, it is interesting to study the effects of orientifolds \cite{MS1,H,MS2,MN5}.
Especially we are interested in how the parallelotopes are deformed by orientifolds.

Fifthly, it is interesting to observe a similarity in degeneracies between the Painlev\'e equations \cite{KNY} and our three-dimensional parallelotopes (figure \ref{cuboctadegenerate}).
For the Painlev\'e equation, after taking the degenerate limit to transform the sixth one into the fifth one, we can either move to the fourth one or the third one, where both degenerate to the second one and furthermore to the first one.
For parallelotopes, after contracting the truncated octahedron into the elongated rhombic dodecahedron, we can either contract it into the cube or the rhombic dodecahedron, where both degenerate to lower dimensions.
The similarity is natural since both of them are characterized by partitions of $4$.
\begin{align}
1+1+1+1\to 2+1+1\;\begin{matrix}\rotatebox[origin=c]{30}{$\to$}\;3+1\;\rotatebox[origin=c]{330}{$\to$}\\[4pt]\rotatebox[origin=c]{330}{$\to$}\;2+2\;\rotatebox[origin=c]{30}{$\to$}\end{matrix}\;4
\end{align}
It is important to elaborate the relations especially from integrabilities \cite{GHM2,G,BGT,JT,2PT,2DTL,BGKNT}.

Sixthly, we have studied the relations between duality cascades and parallelotopes mainly having the setup of three-dimensional supersymmetric Chern-Simons theories in mind.
It is important to see how these relations are generalized to other setups.
Interestingly, relations between duality cascades and zonotopes were studied previously in \cite{EF} for quiver gauge theories obtained from brane tilings \cite{HaKe,FHKVW}.
We would like to clarify relations to parallelotopes also for the setup of brane tilings.

\section*{Acknowledgements}

We are grateful to Naruhiko Aizawa, Hideyuki Ishi, Naotaka Kubo, Zhanna Kuznetsova, Kazunobu Maruyoshi, Kazunobu Matsumura, Tomoki Nakanishi, Takahiro Nishinaka, Tomoki Nosaka, Jaewon Song, Francesco Toppan, Yasuhiko Yamada, Satoshi Yamaguchi and especially Shigeki Sugimoto for valuable discussions and comments.
The work of T.F. and S.M. is supported respectively by Grant-in-Aid for JSPS Fellows \#20J15045 and Grant-in-Aid for Scientific Research (C) \#19K03829, \#22K03598.
S.M.\ would like to thank Yukawa Institute for Theoretical Physics at Kyoto University for warm hospitality.

\end{document}